\documentclass[11pt]{article}
\pdfoutput=1
\usepackage{graphicx,epsfig}
\usepackage{color}
\usepackage[colorlinks,citecolor=red,linkcolor=blue]{hyperref}
\usepackage{slashed}
\usepackage[left=2.5cm, right=2.5cm, top=2cm]{geometry}

\usepackage{epsfig,float,amsmath,amsfonts,amssymb,graphicx,bm,bbm,array,dcolumn}
\usepackage{subfigure}

\newcommand{\Lagr}{\mathcal{L}}

\DeclareMathOperator{\im}{Im}

\newcommand{\Order}{\mathcal{O}}

\def\beq{\begin{equation}}
\def\eeq{\end{equation}}
\def\bea{\begin{eqnarray}}
\def\eea{\end{eqnarray}}
\def\bmat{\begin{pmatrix}}
\def\emat{\end{pmatrix}}
\newcommand{ \slashchar }[1]{\setbox0=\hbox{$#1$}   % set a box for #1
   \dimen0=\wd0                                     % and get its size
   \setbox1=\hbox{/} \dimen1=\wd1                   % get size of /
   \ifdim\dimen0>\dimen1                            % #1 is bigger
      \rlap{\hbox to \dimen0{\hfil/\hfil}}          % so center / in box
      #1                                            % and print #1
   \else                                            % / is bigger
      \rlap{\hbox to \dimen1{\hfil$#1$\hfil}}       % so center #1
      /                                             % and print / 
   \fi}                                             %
\def\etmiss{\slashchar{E}_{T}}
\def\to{\rightarrow}

\graphicspath{{./plots/}}

\begin{document}

\title{
  %\normalsize{\bf
  The 750~GeV diphoton excess in a two Higgs doublet model \\
  and a singlet scalar model, with vector-like fermions, \\
  unitarity constraints, and dark matter implications
}

\author{
%  \normalsize{
%
Shrihari~Gopalakrishna~\thanks{shri@imsc.res.in}~, 
Tuhin Subhra Mukherjee~\thanks{tuhin@imsc.res.in}~,  \\
\small{The Institute of Mathematical Sciences (IMSc),}  
\small{C.I.T Campus, Taramani, Chennai 600113, India.}
}

\maketitle

%%%%%%%%%%%%%%%%%%%%%%%%%%%%%%%%%%%%%%%%%%%%%%%%%%%%%%%%%%%%%%%%%%
\begin{abstract}
  We explore the possibility of a beyond the standard model scalar ($\phi$) as a possible explanation of the 
  diphoton resonance at 750~GeV invariant mass reported by the ATLAS and CMS collaborations at the large hadron collider (LHC).  
  We first present in a model-independent way the scalar-gluon-gluon and scalar-photon-photon effective couplings needed for obtaining 
  the required diphoton cross-section at the LHC for different total widths.
  We investigate here two new-physics possibilities that can generate these effective couplings, namely,  
  (i) the 2-Higgs-doublet model (2HDM) in the alignment limit, and (ii) a singlet scalar,
  with vector-like fermions added and playing a crucial role in generating the effective couplings.
  We present the regions of model parameter space which are allowed by direct LHC and perturbative unitarity constraints, 
  and that give the required diphoton cross-section at the LHC for various total widths.
  In the singlet case, we include the possibility that $\phi$ decays into a pair of neutral stable vector-like fermions that could be dark matter.
  We find regions of parameter-space of the singlet model that gives the required diphoton rate, have the correct dark matter relic-density,
  have dark matter direct-detection rates compatible with current direct-detection experiments, and satisfy LHC bounds and perturbative unitarity constraints.
    
\end{abstract}

\vfill\eject

\tableofcontents

\vfill\eject

%%%%%%%%%%%%%%%%%%%%%%%%%%%%%%%%%%%%%%%%%%%%%%%%%%%%%%%%%%%%%%%%%
\section{Introduction}
\label{Intro.SEC}

The ATLAS and CMS collaborations at the LHC have recently reported an excess of diphoton events at an invariant mass of about $750$~GeV~\cite{ATLAS-750GeVExcess,CMS-750GeVExcess}.
The observations indicate that the cross-section ($\sigma_\phi$) times branching ratio ($BR$)
for the process $pp \to \gamma \gamma$ at a diphoton invariant mass of $750~$GeV is 
about $10\pm 3~$fb~\cite{ATLAS-750GeVExcess} at ATLAS, and  
is $3.75 \pm 1.5$~fb~\cite{CMS-750GeVExcess} at CMS with 8~TeV and 13~TeV data combined. 
The ATLAS best-fit to explain this excess is a resonance with quite a large total width ($\Gamma_\phi \approx 45~$GeV), 
while the CMS best-fit is for a much narrower resonance ($\Gamma_\phi \approx 0.1$~GeV).
There is no statistically significant excess in other channels at this invariant mass. 
We wait to see at the upcoming LHC run if this excess strengthens in significance, a better measurement of the width
is obtained, and if other channels also show excesses. 
Meanwhile, we entertain here the possibility that this excess is due to a new beyond the standard model (BSM) scalar ($\phi$). % and analyze what this implies.

We first present in a model-independent manner, the values the
gluon-gluon-scalar ($\phi gg$) and gamma-gamma-scalar ($\phi\gamma\gamma$)
effective couplings should take to explain the excess, for various total widths and diphoton cross-section values.
If the width of the resonance is large $\Gamma_\phi \gtrsim 1~$GeV, getting the required diphoton rate 
necessitates rather large couplings of the $\phi$ to other states.
Whether such large couplings can violate perturbative unitarity constraints is something we investigate.

We then consider different models which contribute to these effective couplings. 
The models we consider are very generic and can be embedded into various BSM frameworks. 
In particular, the two possibilities we consider are that
(i) $\phi$ is in the doublet representation of $SU(2)$ gauge group of the standard model (SM), and
(ii) in the singlet representation of $SU(2)$.
In addition we introduce vector-like fermions (VLF), namely vector-like leptons (VLL) and vector-like quarks (VLQ) coupled to the $\phi$.
The doublet or singlet scalars coupled to SM fermions (SMF) and/or VLFs that we deal with here
can be thought of as extracts from various BSM constructions that are relevant to explain the diphoton rate.

In the case of the doublet, we consider $\phi$ as the CP-even and CP-odd scalars of a two-Higgs-doublet model (2HDM), 
and we include Type-I, Type-II and Type-X possibilities for SMF couplings.
We compute the SMF and VLF contributions to the $\phi gg$ and $\phi\gamma\gamma$ effective couplings at 1-loop.
A colored fermion can contribute to the $\phi gg$ and $\phi\gamma\gamma$ couplings, 
while a color singlet charged fermion can contribute to the latter.
Also, if the fermion mass is less than $M_\phi/2$, the $\phi$ decays into such fermion pairs contributes to the total width of the $\phi$.

In the case of the singlet $\phi$, if the $\phi$ and the SM Higgs ($h$) are coupled via a cubic or quartic interaction,
the $\phi$ could mix with $h$ after spontaneous symmetry breaking. 
This mixing induces a coupling between a hidden sector that the $\phi$ is a part of and the visible sector (SM), 
If the hidden sector contains a neutral and stable singlet VLF $\psi$, this could be dark matter. 
In this case, $\phi \to \psi\psi$ decays contribute to $\Gamma_\phi$ if $M_\psi < M_\phi/2$, 
$\phi$ exchange controls the relic-density via the self-annihilation channel $\psi\psi \to SM$ in the early universe, 
and $\phi$ exchange can mediate the interaction of the dark matter with a nucleon leading to direct detection prospects.
Interestingly, in the limit of this mixing going to zero, the visible and hidden sectors do not decouple if VLFs are present, 
since the effective $\phi gg$ and $\phi \gamma\gamma$ couplings induced by VLFs remain as couplings between the two sectors.  
This sets the relic density, and also leads to direct-detection. 
We investigate these aspects in this work.

If the width of the resonance is large $\Gamma_\phi \gtrsim 1~$GeV, getting the required diphoton rate 
necessitates rather large couplings of the $\phi$ to VLFs.
Whether such large couplings can violate perturbative unitarity constraints is something we investigate.

Next, we make contact with other works in the literature that have overlap with our work.
We perform a model-independent analysis and present the sizes of $\phi g g$ and $\phi\gamma\gamma$ effective couplings required
to explain the diphoton excess for various $\phi$ total widths.
Similar model-independent analysis is done for example in Refs.~\cite{Franceschini:2015kwy,Gupta:2015zzs}
but for a fixed value of total $\phi$ width.
Refs.~\cite{Djouadi:2016eyy}-\cite{Angelescu:2015uiz} interpret the resonance as scalars of the 2HDM type-I and 2HDM type-II with VLFs.
In addition to analyzing the type-I and type-II couplings, in this work we also include the 2HDM type-X, 
take into account the limits from the 8~TeV LHC $\phi \to tt$ and $\phi \to \tau \tau$ channel results for all these types, and 
find perturbative unitarity constraints from $\phi \phi \to \phi \phi$ and $\psi \psi \to \psi \psi$ channels. 
These have not been considered in the literature so far.
Refs.~\cite{Franceschini:2015kwy,Gupta:2015zzs,Djouadi:2016eyy} and \cite{DiChiara:2015vdm}-\cite{Backovic:2015fnp} include interpretation of the resonance as a singlet scalar coupled to VLFs.
Refs.~\cite{Bhattacharya:2016lyg,D'Eramo:2016mgv,Chen:2016sck,DiChiara:2016dez,Mambrini:2015wyu,Ko:2016wce,Redi:2016kip,Backovic:2015fnp} additionally
discuss the dark matter implications of a neutral VLF coupled to the scalar.
Ref.~\cite{McDermott:2015sck} considers a singlet $\phi$ coupled to an SU(2) singlet VLL of EM charge 1
and an SU(2) singlet VLQ with various EM charges;
$h \leftrightarrow \phi$ mixing and $\phi h h$ coupling were not included, which we do.
We also explore the possibility of the VLL having zero EM charge and
it being a dark matter candidate. 
Such a study has also been carried out in 
Ref.~\cite{Bhattacharya:2016lyg} but with only VLLs and no VLQs.
Our work includes VLQs. 
Furthermore, they do not allow $\phi$ decays to the VLLs as we do here to obtain a large $\phi$ width.
In our work we include the contribution of the $\phi gg$ effective coupling induced by VLQs to the direct detection process mediated by the $\phi$. 
This contribution is present even when the the Higgs-singlet mixing is either very small or not present.
This is an important contribution in our case, which is usually not included in the literature. 
Usually in the literature, only the $h$ contribution is included, which for very small Higgs-singlet mixing is a small contribution.
Ref.~\cite{Ge:2016xcq} does include this contribution, which is sub dominant in their case with the main contribution being due to the Higgs. 
Furthermore, in their case the dark matter is not a VLL as in ours, 
and it does not discuss the constraints from the 8~TeV LHC $\phi \to hh$ result, which we include.
In the singlet scalar model, we find constraints on the parameter space
from the requirement of perturbative unitarity in the $\phi \phi \to \phi \phi$ and $\psi \psi \to \psi \psi$ channels. 
This has not been considered in the literature.

The rest of the paper is organized as follows.
In Sec.~\ref{ModInd.SEC} we present a model-independent analysis for a general 750~GeV scalar coupled to VLFs.
In Sec.~\ref{MatchData.SEC} we present the values of $\phi gg$ and $\phi \gamma \gamma$ effective couplings
required to explain the observed $\sigma(pp \to \gamma \gamma)$ for different total $\phi$ width.
In Sec.~\ref{constr.SEC} and Sec.~\ref{utrtyCon.SEC} we discuss the 8~TeV LHC constraints
and perturbative unitarity constraints respectively.
In Sec.~\ref{MODELS.SEC} we analyze few specific models as mentioned in the introduction
and present regions of the parameter space which can generate the observed $\sigma(pp \to \gamma \gamma)$ cross section,
while being consistent with the 8~TeV LHC constraints and the perturbative unitarity constraints.
In Sec.~\ref{2HDM.SEC} we analyze the 2HDM type-I, type-II and type-X, coupled to VLFs.
In Sec.~\ref{singPhiHidDM.SEC} we analyze the singlet scalar model coupled to VLFs.
We also discuss the dark matter implications of the singlet scalar model in Sec.~\ref{HidSec.SEC}.
In Sec.~\ref{DisCon.SEC} we offer our conclusions, and point out some promising signals
to look for at the upcoming LHC to ascertain if any of the the models we
discuss are realized in nature.
In App.~\ref{hidRelDen.APP} we present the relevant formulas for the dark matter relic density calculation in the singlet scalar model.
%%%%%%%%%%%%%%%%%%%%%%%%%%%%%%%%%%%%%%%%%%%%%%%%%%%%%%%%%%%%%%%%%
\section{Model-independent analysis}
\label{ModInd.SEC}

We explore the possibility that the $750~$GeV resonance a scalar $\phi$ with $M_\phi = 750~$GeV.
We start by effectively parameterizing the fermion interactions with $\phi$ as
\beq
\Lagr \supset
- \frac{y_\psi}{\sqrt{2}} \phi \bar\psi \psi - \frac{y_f}{\sqrt{2}} \bar f_L H f_R + h.c. \ ,
\label{LmodInd.EQ}
\eeq
where $H$ denotes the SM Higgs doublet containing the physical Higgs boson $h$ with $m_h = 125$~GeV,
$\phi$ denotes the new scalar with $M_\phi = 750~$GeV, $\psi$ denotes new vector-like fermions (VLF), and $f_{L,R}$ denotes SM fermions (SMF).
In this section we perform a model-independent analysis using effective operators relevant to the diphoton excess. 
In Sec.~\ref{MODELS.SEC} we consider models in which $\phi$ is either an SU(2) singlet or is embedded in an SU(2) doublet.
The $\psi$ represents a set of either colored vector-like fermions, or color-singlet vector-like fermions that are EM charged or neutral.

%%%%%%%%%%%%%%%%
%%%%%%%%%%%%%%%%%%%%%%%%%%%%%%%%%%%%%%%%
%\medskip
%\noindent \underline
\subsection{Matching to the diphoton data}
\label{MatchData.SEC}
In Sec.~\ref{Intro.SEC} we quoted the ATLAS and CMS best-fit diphoton rates and total widths. 
Here we determine the sizes of the $\phi g g$ and $\phi \gamma\gamma$ effective couplings needed to explain the excess. 
We work in the narrow width approximation (NWA) in which we can write
$\sigma (pp \to \phi \to \gamma \gamma) \approx \sigma (p p \to \phi) * BR(\phi \to \gamma \gamma) \equiv \sigma_\phi * BR_{\gamma\gamma} $ with
$BR(\phi \to \gamma \gamma) \equiv \Gamma(\phi \to \gamma \gamma)/\Gamma_\phi$ where $\Gamma_\phi$ is the total width of the $\phi$.
We consider here $\phi$ production via the gluon-fusion channel.  
Rather than compute $\sigma (gg \to \phi)$ ourselves, we relate it to the SM-like Higgs production c.s. at this mass and make use of the vast literature on this by writing
\beq
\sigma(gg\to \phi) = \sigma(gg\to \phi_{\rm SM}) \frac{\Gamma(\phi \to g g)}{\Gamma(\phi_{\rm SM} \to gg)} \ ,
\label{pp2phiGmgg}
\eeq
where $\phi_{\rm SM}$ denotes a SM-like Higgs with mass $M_{\phi_{\rm SM}} = 750~$GeV for which $\sigma(gg\to \phi_{\rm SM}) = 0.7 \pm 0.2~$pb~\cite{Baglio:2010ae} at the 14~TeV LHC.
We scale this to $\sqrt{s}=13~$TeV and take $\sigma(gg\to \phi_{\rm SM}) = 0.6 \pm 0.2~$pb.
For example, if the new state $\phi$ couples to gluons with the same effective coupling strength as the $\phi_{\rm SM}$,
in order to get the required $\sigma_\phi * BR_{\gamma\gamma}$ to match the excess, we need $BR(\phi \to \gamma\gamma) \sim 10^{-3}$.
As can be inferred from Eq.~(\ref{pp2phiGmgg}) and detailed in Ref.~\cite{Gopalakrishna:2015wwa}, 
a colored fermion (quark) coupled to $\phi$ via a Yukawa coupling $y'_f/\sqrt{2}$, 
gives a contribution to
\beq
\sigma(gg \to \phi) = \sigma(gg\to \phi_{\rm SM}) \left|\sum_{f} \frac{ y_f}{\hat{y}_t} \frac{F_{1/2}(\tau_f)}{F_{1/2}(\tau_t)} \frac{m_t}{M_{f}} \right|^2 \ ,
\label{gg2phiySq}
\eeq
where $\hat{y}_t = \sqrt{2} m_t/v$ is the $\hat{h}tt$ Yukawa coupling, $\tau_f \equiv M^2_\phi/(4 m^2_f)$, and $F_{1/2}$ is defined in Ref.~\cite{Gopalakrishna:2015wwa}.
The sum over $f$ in the numerator includes the top-quark contribution. 

Defining $\alpha \equiv \Gamma_\phi / M_\phi$ we see that $\alpha \approx 0.06$ for ATLAS best-fit $\Gamma_\phi$, while $\alpha \approx 1.4\times 10^{-4}$ for CMS best-fit $\Gamma_\phi$.
We await further confirmation from ATLAS and CMS as to what the true $\Gamma_\phi$ is; 
meanwhile, in this work we vary $\alpha$ to cover this entire range of $\Gamma_\phi$. 
Including all the decay modes of the $\phi$, we can write for the total width of the $\phi$
\beq
\Gamma_\phi \equiv \frac{\kappa_{\Gamma}^2}{16 \pi} M_\phi \ ,
\label{GmphiKap}
\eeq
which defines $\kappa_{\Gamma}^2$ to include all couplings relevant to $\phi$ decay, and phase-space factors as appropriate.
We thus infer that $\kappa_\Gamma^2 = 16 \pi \alpha$, which implies that for $\alpha = 0.06$, we need $\kappa_\Gamma^2 = 3$ (ATLAS best-fit),
and for $\alpha = 1.4\times 10^{-4}$, we need $\kappa_\Gamma^2 = 7\times 10^{-3}$ (CMS best-fit). 
For example, for the decay into colored fermions much lighter than $M_\phi$, coupled via a Yukawa coupling $y_0/\sqrt{2}$,
we have $\kappa_{\Gamma}^2 = N_c y_{\phi f_o f_o}^2$ and for $N_c = 3$ we need $y_{\phi f_o f_o} = 1$, if $\Gamma_\phi$ is as claimed by ATLAS.
This large of a width requires that $\phi$ couples with an $\Order (1)$ coupling strength to some state that it decays to.
Generally speaking, if we take $\alpha = 0.06$, the large total width suppresses the BR into loop suppressed decay modes such as
$\phi\to\gamma\gamma$, and it will be nontrivial to get $BR(\phi \to \gamma\gamma) \sim 10^{-3}$ in any new physics model as required to explain the excess. 
We will study in Sec~\ref{MODELS.SEC} to what extent we can achieve this and its implications.

\begin{figure}
\centering
\includegraphics[width=0.32\textwidth]{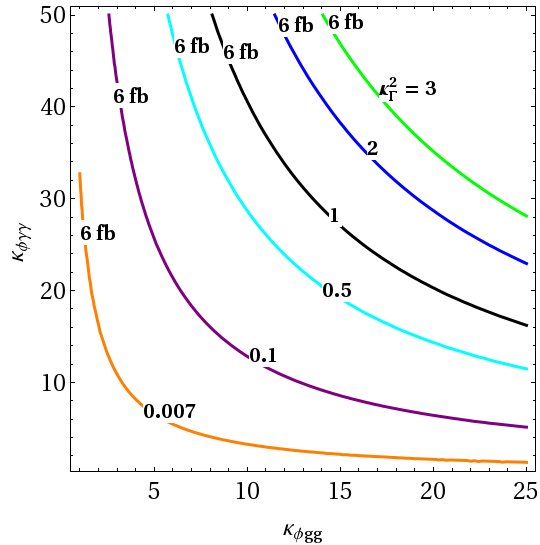}
\includegraphics[width=0.32\textwidth]{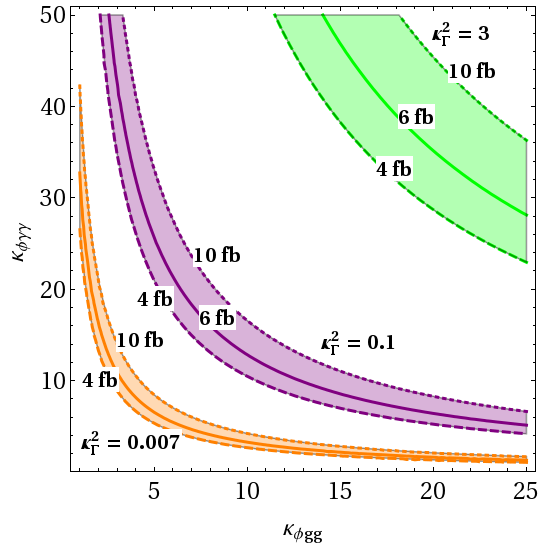}
\caption{For various $\kappa^2_\Gamma$ shown, the $\kappa_{\phi gg}$ and $\kappa_{\phi \gamma\gamma}$ required for $\sigma_\phi * BR_{\gamma\gamma} = 6~$fb (left),
  the regions $4 \leq \sigma_\phi * BR_{\gamma\gamma} \leq 10$ around $\kappa^2_\Gamma = 3$ (green), $0.1$ (purple), $0.007$ (orange) curves (right).}
\label{kggkaaAl}
\end{figure}

Before dealing with specific models, we find the sizes of effective couplings that are needed to explain this excess.
We follow the notation and effective coupling definitions of Ref.~\cite{Gopalakrishna:2015wwa} for the $\phi$-gluon-gluon and $\phi$-photon-photon effective couplings,
$\kappa_{\phi gg}$ and $\kappa_{\phi \gamma\gamma}$ respectively.
The color factor in $\kappa^{ab}_{\phi g g}$ is $C_{ab} = (1/2)\delta_{ab}$, where $a,b=\{1,...,8\}$ are the adjoint color indices. 
In the plots below and in Ref.~\cite{Gopalakrishna:2015wwa} App.B, we include this factor of (1/2) in the $\kappa_{\phi g g}$ and suppress the color indices. 
Computing a decay rate or cross-section by summing over $a,b$ gives
$\sum_{a,b} |C_{ab}|^2 = 8 (1/2)^2 = 2$ resulting in a color factor of 2.
From Eqs.~(\ref{pp2phiGmgg})~and~(\ref{GmphiKap}) we can write
\beq
\label{sigmaphiBrGm}
\sigma_\phi * BR_{\gamma\gamma} = \left[ \sigma(gg\to \phi_{\rm SM}) \frac{ \kappa_{\phi gg}^2 }{ \kappa_{\hat{h}gg}^2 } \right] * \left[ \frac{1}{4} \left(\frac{\kappa_{\phi \gamma\gamma}}{16\pi^2 M}\right)^2 \frac{M_\phi^2}{\kappa_\Gamma^2} \right] \ ,
\eeq
where $M$ is a reference mass-scale which we set to $1~$TeV.
Expression for the $\Gamma(\phi \to XX)$ can be found for example in Refs.~\cite{Gopalakrishna:2015wwa,Gunion:1989we}. 
We find that $\kappa_{\hat{h}gg} = 10$. 
In Fig.~\ref{kggkaaAl} we show for various $\kappa^2_{\Gamma}$ the $\kappa_{\phi gg}$ and $\kappa_{\phi \gamma\gamma}$ required for $\sigma_\phi * BR_{\gamma\gamma} = 6~$fb,
taking this value as a representative diphoton cross section that explains the excess.
We also show in Fig.~\ref{kggkaaAl}
a band around $\kappa^2_\Gamma = {3, 0.1, 0.007}$, three representative total width values.

Model-independently, we can define an effective $\phi h h$ coupling as
\beq
{\cal L} \supset - \frac{\kappa_{\phi h h} M_\phi}{2 \sqrt{2}} \phi h^2 \ .
\label{phihhDefn.EQ}    
\eeq
This term leads to the $\phi \to h h$ decay, which as we see below is constrained at the LHC. 
We find that the $\kappa_{\phi h h}$ contributes to $\kappa_\Gamma^2$ an amount $ (\kappa_{\phi h h}^2/4)\sqrt{1-4 m_h^2/M_\phi^2}$.

%%%%%%%%%%%%%%%%%%%%%%%%%%%%%%%%%%%%%%%%%%%%%%%%%%%
\subsection{LHC constraints}
\label{constr.SEC}

If the $\phi t\bar t$ and $\phi \tau\bar\tau$ couplings are nonzero, 
$\phi$ decays to $t\bar t$ and $\tau\bar \tau$ also.
Since there are no reported excesses in these channels, there could be nontrivial constraints on the models from these channels.
We discuss these constraints next.

In Ref.~\cite{Gopalakrishna:2015wwa} Fig.~2, we show constraints on the $\kappa_{\phi gg}$ from the 8~TeV LHC exclusion limits.
To summarize this, for $BR_{t\bar t} = 1$, the constraint from the $t\bar t$ channel is $\kappa_{\phi g g} < 20$,
and for $BR_{\tau\bar\tau} = 1$, the constraint from the $\tau\bar\tau$ channel is $\kappa_{\phi g g} < 4$.
Of course, in a particular model, these BRs can be significantly smaller than $1$, particularly $BR_{\tau\tau}$, and the limits can be correspondingly weaker.  
For a SM-like theory with only the Higgs mass set at $750$~GeV, 
we have $\kappa_{\phi_{\rm SM} g g} = 10$ with $\sigma(pp\to \phi_{\rm SM}) \approx 100~$fb at the 8~TeV LHC
due mainly to the top contribution.
From the $\kappa_{\phi gg}$ expressions in Eq.~(B.1) of Ref.~\cite{Gopalakrishna:2015wwa} and with
$BR_i = \kappa_{i}^2/\kappa_\Gamma^2$, 
we derive the bound
\beq
\left|\sum_{Q} \frac{ y_{\phi QQ}}{\hat{y}_t} \frac{F_{1/2}(\tau_Q)}{F_{1/2}(\tau_t)} \frac{m_t}{M_{Q}} \right|^2 \frac{\kappa_{i}^2}{\kappa_\Gamma^2}  < \left(\frac{\kappa_{\phi g g\, (i)}^{\rm max}}{\kappa_{\phi_{\rm SM} g g}}\right)^2 \ ,
\eeq
where the sum over $Q$ includes the top-quark contribution plus any new colored vector-like fermions present in the $\phi g g$ loop,
$\hat{y}_t \approx 1$ is the SM top Yukawa coupling (we ignore the effect of running this to the scale $\mu = M_\phi$),
and $\kappa_t^2 = N_c y_{\phi t t}^2 (1-4 r_t)^{n/2}$, $\kappa_\tau^2 = y_{\phi \tau \tau}^2 (1-4 r_\tau)^{n/2}$
with $n=3$ for a CP-even $\phi$ and $n=1$ for a CP-odd $\phi$, and $r_f \equiv M_f^2/M_\phi^2$.
The index $(i)$ runs over various channels $\{t\bar t, \tau\bar\tau, hh, gg, ...\}$ i.e. $(i) = \{t,\tau, h, g \}$, 
and we have $\kappa_{\phi g g\, (t)}^{\rm max} = 20$, $\kappa_{\phi g g\, (\tau)}^{\rm max} = 4$ (corresponding to $BR_i = 1$)  as
derived in Ref.~\cite{Gopalakrishna:2015wwa}.
The LHC upper limit on the $hh$ channel $\sigma * BR$ at a mass of $750~$GeV is about $30~$fb~\cite{Aad:2015xja}, 
which translates into $\kappa_{\phi g g\, (h)}^{\rm max} = 3.3$. 
The LHC upper limit on the dijet channel at a mass of $750~$GeV is about $2~$pb~\cite{CMS:2015neg, Aad:2014aqa}, 
and for the sizes of cross-section and dijet BR we are dealing with here, this will not be a nontrivial constraint. 

%%%%%%%%%%%%%%%%%%%%%%%%%%%%%%%%%%%%%%%%%%%%
Generically, in new physics models there are shifts in the $h$ couplings to SM states, which are constrained by the LHC data (see for example Ref.~\cite{Ellis:2014dza}).
In the models we consider below, we pay attention to this constraint and ensure that these do not violate the constraints. 

The models we discuss below also contain vector-like fermions, and there are direct limits on them also from the LHC, which have to be obeyed.
Preventing a stable cosmological colored relic implies that they have to be allowed to mix with SM fermions.
Allowing only mixing to third generation SM quarks is sufficient and is relatively safer with respect to FCNC constraints. 
We assume that there are small off diagonal mass mixing terms $\delta m$ to third-generation quarks such that $\delta m/M_{VL} \lesssim 0.1$ to third generation SM quarks
but big enough such that the VLQ decays
such that it is not in conflict with cosmological data. 
We summarize next the present LHC lower limits on VLF masses, with the precise limit depending on the BRs. 
The lower limit on the $t'$ mass is presently in the range $750-920~$GeV~\cite{ATLAS:2013ima,CMS:2013tda,Khachatryan:2015oba,ATLAS:tp-13TeV-3.2ifb,Aad:2015kqa},
and on the $b'$ mass in the range $740-900$~GeV~\cite{Aad:2015kqa,Aad:2015mba,Khachatryan:2015gza}.
For a long-lived VLQ with life-times in the range $10^{-7} - 10^5$~s, the bound is looser with $M_Q \gtrsim 525~$GeV being allowed~\cite{Khachatryan:2015jha,Aad:2013gva}. 
\footnote{
It may be possible to weaken the VLQ mass bound somewhat by allowing $t' \to t \phi^\prime$ and/or $b' \to b \phi^\prime$ decays, where $\phi^\prime$ is an SU(2) singlet and will lead to missing energy at the LHC. 
This for example can be achieved by introducing the operators $U \phi' t^c$ or $B \phi' b^c$ where the $U$ and $B$ are the charge $2/3$ and $-1/3$ SU(2) singlet VLFs,
$t^c$ and $b^c$ are SM SU(2) singlet fermions.
Due to the new decay mode, the usual assumption that the BRs into the SM final states ($bW,tZ,th$ for the $t'$ for instance) sum to one fails, and the limits have to be reanalyzed.
The BRs into the SM final states are decreased and since the new mode has substantially larger SM irreducible SM $t\bar t + \etmiss$ (or $b\bar b + \etmiss$) backgrounds,
the VLQ lower limits should be weaker.
A detailed investigation of the implications of this proposal
is beyond the scope of this work.
}
The lower limit on VLL masses is presently $ \lesssim 100~$GeV if it decays only into a $\tau$,
and about $300~$GeV ($450~$GeV) for a singlet (doublet) that decays into $e,\mu$~\cite{Falkowski:2013jya}.

%%%%%%%%%%%%%%%%%%%%%%%%%%%%%%%%%%%%%%%%%%%%%%%%%%%%%%%%%%%%%%%%%%%
\subsection{Unitarity constraint}
\label{utrtyCon.SEC}

\begin{figure}
\centering
\includegraphics[width=0.45\textwidth]{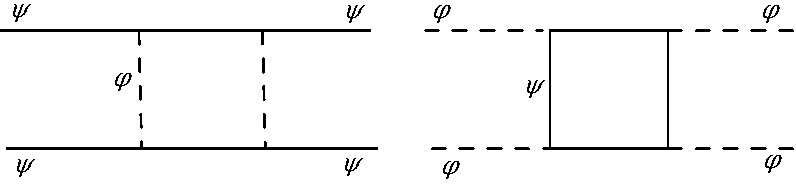}
\caption{1-loop box diagrams contributing to $\psi \psi \to \psi \psi$ (left) and $\phi \phi \to \phi \phi$ (left) processes.
}
\label{unitrb}
\end{figure}

%%%%%%%%%%%%%%%%%%%%%%%%%%%%%%%%%%%%%%%
If the large $\Gamma_\phi$ is
due to a large decay width into some fermion $\psi$ coupled as in Eq.~(\ref{LmodInd.EQ})
and with a large $y_\psi$, there is a limit to how large $y_\psi$ can be if perturbative unitarity is to be maintained.
This limit can be worked out by considering, for example, $\psi\psi \to \psi \psi $ scattering.
The tree level contribution leads to a very loose bound, and we therefore consider the 1-loop box diagram shown in Fig.~\ref{unitrb}.
Expanding the amplitude ${\cal M}$ in partial waves as
\beq
{\cal M}(\cos\theta) = 16 \pi \sum_l (2l + 1) a_l P_l(\cos\theta)\ ,
\eeq
a necessary condition for unitarity is $\im{a_l} \leq 1$~\cite{Agashe:2014kda}. 
Using the optical theorem, we can compute 
the $\im({\cal M}(\psi\psi\to\psi\psi))$ for forward scattering in terms of the cross section $\sigma(\psi\psi \to \phi \phi)$ with the precise relation given by (see for example Ref.~\cite{Peskin:1995ev})
\beq
\im\left\{{\cal M}\left(\psi(p_1) \psi(p_2) \to \psi(p_1) \psi(p_2)\right)\right\} = 2 E_{cm} p_{cm} \, \sigma \left(\psi(p_1) \psi(p_2) \to \phi(k_1) \phi(k_2) \right) \ . 
\eeq
For $\sigma$ we compute the spin-averaged cross section for $\psi \psi \to \phi \phi$ scattering as
\beq
\frac{d\sigma}{d\cos\theta} = \frac{y_\psi^4}{128\pi s} \tilde{f}(\cos\theta) \ , \quad {\rm where}\ 
\tilde{f}(\cos\theta) = \frac{(1-2r_{\phi E})^{3/2}(1-\cos^2\theta)}{2(1-r_{\phi E} - \sqrt{1-2r_{\phi E}} \cos\theta)^2} \ ,  
\eeq
where we have ignored the fermion mass, and $r_{\phi E} \equiv M_\phi^2/(2 E_\psi^2)$.
Integrating this over $\cos\theta \subset (-1,1)$, we obtain for $E_\psi \gtrsim M_\phi$ the approximate equality
$\sigma(\psi\psi \to \phi \phi) \approx y_\psi^4/(128\pi s)$.
To get a conservative bound, we assume that the $\psi\psi \to \psi\psi$ amplitude is saturated by the $l=0$ partial-wave, and obtain
$ \im{a_0} = y_\psi^4/(4\times (16\pi)^2) < 1$, i.e. $y_\psi < 10$ as the unitarity bound .  
The related process $\phi\phi \to \phi\phi$ also leads to a similar bound, but the amplitude is enhanced by $N_c$ for a colored fermion in the 1-loop amplitude
of Fig.~\ref{unitrb}, and leads to a bound $\sum_f y_f (N_c^f)^{1/4} \lesssim 10$ where $N_c^f = 3$ for a colored fermion
(and $N_c^f = 1$ for an uncolored fermion),
and the sum is over all fermions that contribute in the loop.

%%%%%%%%%%%%%%%%%%%%%%%%%%%%%%%%%%%%%%%%%%%%%%%%%%%%%%%%%%%%%%%%%
\section{Models}
\label{MODELS.SEC}

In this section we explore in turn two classes of models, namely $\phi$ is in an SU(2) doublet, 
and after that $\phi$ being an SU(2) singlet. We identify regions of parameter space 
which are safe with respect to the unitarity and LHC constraints, 
and which give the required diphoton rate. 

%%%%%%%%%%%%%%%%%%%%%%%%%%%%%%
\subsection{2-Higgs-Doublet Model}
\label{2HDM.SEC}
We consider here the 2HDM
and briefly summarize below aspects of the 2HDM relevant for our work; 
for a comprehensive review of the 2HDM see Ref.~\cite{Branco:2011iw}.
Our notation here is as defined in Ref.~\cite{Gopalakrishna:2015wwa}.
The 2HDM contains two scalar $SU(2)$ doublets $\Phi_1$ and $\Phi_2$ both of which we take to have hypercharge $Y=1/2$. 
We do not show the potential explicitly here, and
the electroweak vacuum is obtained by minimizing the potential with respect to $\Phi_1$ and $\Phi_2$.
The neutral components of $\Phi_1, \Phi_2$
get vacuum expectation values denoted by $v_1,v_2$ respectively,
with $v^2 = v_1^2 + v^2_2 = (246~\text{GeV})^2$.
We take $\tan\beta = v_2/v_1 $.
After electroweak symmetry breaking different components of $\Phi_1$
mixes with corresponding components of $\Phi_2$ giving rise to five physical states in unitary gauge, which are, 
the two CP-even neutral scalars $h,H$, the CP-odd neutral scalar $A$, and the charged scalar $H^\pm$.
$A$ is a linear combination of the CP-odd scalars in $\Phi_1, \Phi_2$ with the the mixing angle given by $\tan \beta$, 
the other linear combination being the Goldstone boson that is not in the physical spectrum in unitary gauge. 
$h,H$ are linear combinations of the two CP-even scalars of $\Phi_1$ and $\Phi_2$ with the mixing angle denoted 
as usual by $\alpha$.
We will identify the $h$ with the 125 GeV Higgs.
In the so called ''2HDM alignment limit''~\cite{Gopalakrishna:2015wwa,Bernon:2015qea,Bhattacharyya:2015nca} given by $\beta - \alpha = \pi/2$,
the Yukawa couplings and the gauge couplings of the $h$ become identical to those of the SM Higgs.
In this work we will always work in the alignment limit.
If an SMF couples to both $\Phi_1$ and $\Phi_2$, tree level FCNCs result, severely constraining the model.  
To be safe from this, usually, a $Z_2$ symmetry is imposed on the $\Lagr$
under which $\Phi_1 \to -\Phi_1$ and $\Phi_2 \to \Phi_2$.
The fermion $Z_2$ transformation is fixed depending on which of the $\Phi_1,\Phi_2$ it couples to.
The usual types of couplings well known in the literature that we study here are 
the so called 2HDM type-II, type-X and type-I which we denote as 2HDM-II, 2HDM-X and 2HDM-I.

We take $M_H, M_A=735, 750$~GeV, and find out if the model can explain the diphoton excess, with the $A,H$ contributing. 
Our results presented here do not depend very sensitively on this mass splitting.
It is possible that the ATLAS large width is an apparent effect due to the presence of two narrower Briet-Wigner resonances due to the decays of $A$ and $H$.
The combined line-shape is shown for example in Ref.~\cite{Djouadi:2016eyy}.
Since $H$ and $A$ have opposite CP quantum numbers their contribution in a channel is incoherent,
that is $\sigma*BR_{\gamma \gamma} = \sigma_H*BR(H \to XX) + \sigma_A*BR(A \to XX)$.
We analyze the situation with only the SMF present,
and subsequent to this with the addition of vector-like fermions (VLF), namely colored vector-like quarks (VLQ) and vector-like leptons (VLL).
We present our results only for SM-like VLF hypercharge assignments, and for larger EM charges our results on the diphoton rate can be scaled by $Q_f^4$.
Also, we add only one copy of VLFs but our diphoton results can again be scaled by the number of copies quite easily, although a very mild tightening of the $\phi\phi\to\phi\phi$ unitarity bound results which scales
like the fourth root of the number of copies. 
We explore type-I, type-II, and type-X SMF couplings, but keep the VLF couplings as in type-II;
taking other types for the VLF couplings is also a possibility, which we do not study here, 
for which the diphoton rates may differ by factors of a few.
We draw heavily from the work in Ref.~\cite{Gopalakrishna:2015wwa} which analyzes such a scenario.
The expressions for $\kappa_{\phi gg }$ and $\kappa_{\phi \gamma\gamma}$ are given in App.~B of Ref.~\cite{Gopalakrishna:2015wwa}.

As mentioned in Sec.~\ref{constr.SEC}, the $hff$ couplings are consistent with the SM to the accuracy measured at the LHC.
Although in general the $hff$ couplings are shifted in the 2HDM, remarkably, they coincide with the SM values in the alignment limit, provided the fermion couples only to one of $\Phi_1$ or $\Phi_2$,
which we will assume is the case. 
If the fermion couples to both $\Phi_1$ and $\Phi_2$, the $hff$ coupling shift imposes nontrivial constraints on the model. 
These aspects are explained in detail for example in Ref.~\cite{Gopalakrishna:2015dkt}.

The 8~TeV $hh$ channel constraints discussed in Sec.~\ref{constr.SEC} constrains $\kappa_{\phi hh} \ll 1$. 
For example, for $\kappa_{\phi g g} \approx 10$ leading to $\sigma_\phi \approx 1~$pb, $\kappa_{\phi hh}$ in a particular 2HDM model must be small enough that
$BR_{hh} \lesssim 0.05$.
For example, in the 2HDM little-Higgs model of Ref.~\cite{Gopalakrishna:2015dkt} we have $\kappa_{\phi h h} \propto (M_A^2 - M_{H^\pm}^2)/(v M_\phi) \approx 0.04$.
In the 2HDM models we discuss below, we assume that the 2HDM potential (that we do not specify) is such that $\kappa_{\phi h h}$ obeys this constraint. 

%%%%%%%%%%%%%%%%%%%%%%%%%%%%%%%%%%%%%%%%
\subsubsection{2HDM type-II}
In the 2HDM type-II (2HDM-II) model the up-type SMFs couple only to $\Phi_2$ 
and the down-type SMFs couple only to $\Phi_1$.
The mass of the up and down-type fermions are given by $(y_f v_2/\sqrt{2})$ and $(y_f v_1/\sqrt{2})$ respectively.
The Yukawa couplings of the fermions to $H,A$ are respectively $(y_f \sin\alpha/\sqrt{2})$, $(y_f \cos\beta/\sqrt{2})$
for up-type fermions and $(y_f \cos\alpha/\sqrt{2})$, $(y_f \sin\beta/\sqrt{2})$ for down type fermions.
We can trade the $y_f$ for the fermion masses $m_f$.
As stated earlier, we will take the alignment limit.
Since $y_{\phi tt} \propto \cot \beta$ and $y_{\phi bb} \propto \tan \beta$,
$\kappa^2_\Gamma$ can not be made arbitrarily small in this model;
the minimum occurs at $\tan \beta \simeq 5.7$ corresponding to $\kappa^2_\Gamma = 0.12$ when only $H$ contribute
and $0.24$ when both $H$ and $A$ contribute.
Any value of $\kappa^2_\Gamma$ (other than 0.24) can be realized by two values of $\tan \beta$;
one for $\tan \beta < 5.7$ and the other for $\tan \beta > 5.7$.

\begin{figure}
\centering
\includegraphics[width=0.45\textwidth]{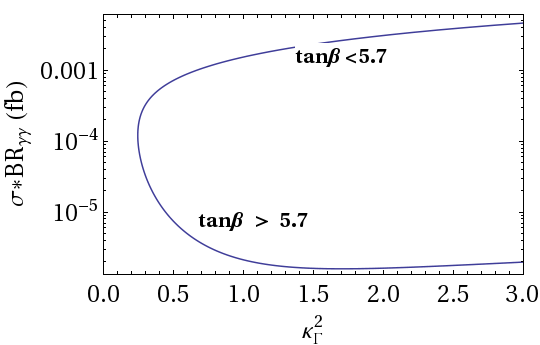}
\caption{In 2HDM-II with SMF only, $\sigma*BR_{\gamma \gamma}$ vs. $\kappa^2_{\Gamma}$ obtained by varying $\tan \beta$ with $M_H,M_A= 735,750$~GeV.
}
\label{sigmabr-2HDM.SMF}
\end{figure}

%%%%%%%%%%%%%%%%%%
\medskip
\noindent \underline{2HDM-II with SMF only}:
If only $H$ is included for illustration, we have $\kappa^2_\Gamma = 3$ and $\sigma*BR_{\gamma \gamma} \simeq 0.002$~fb for $\tan \beta =0.83$.
In reality, the nearly degenerate $H,A$ both contribute to $\sigma*BR_{\gamma \gamma}$, and 
in Fig.~\ref{sigmabr-2HDM.SMF} we show $\sigma*BR_{\gamma \gamma}$ vs. $\kappa^2_\Gamma$
obtained by varying $\tan \beta$ for $M_H, M_A = 735,750$~GeV.
The two branches of $\sigma*BR_{\gamma \gamma}$ in Fig.~\ref{sigmabr-2HDM.SMF} correspond to
two values of $\tan \beta$ that gives the same $\kappa^2_\Gamma$.
The upper branch for which $\tan \beta < 5.7$ has larger cross sections because of the larger contribution from the top.
%

\begin{figure}
\centering
\includegraphics[width=0.45\textwidth]{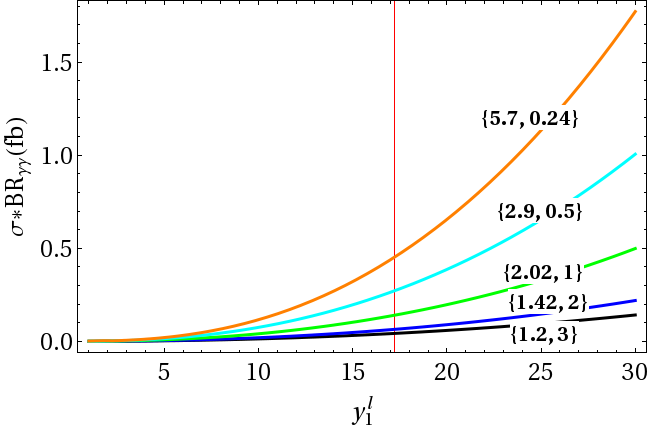}
\includegraphics[width=0.45\textwidth]{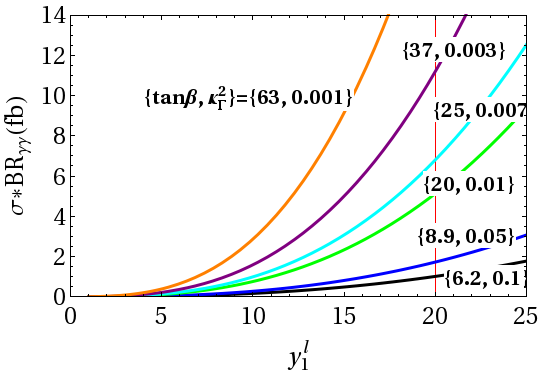}
\caption{In 2HDM-II (left) and 2HDM-I (right) with SMF+VLL,
for $M_H =735~$GeV, $M_A = 750~$GeV and various $\{\tan \beta, \kappa^2_\Gamma\}$ shown, $\sigma*BR_{\gamma \gamma}$ with VLL mass parameters chosen such that
the lighter VLL mass eigenvalue is fixed at $375~$GeV.
The unitarity constraint on $y^l_1$ from $\psi \psi \to \psi \psi$ process is shown by the vertical red line.
}
\label{sigmabr-2HDM.VLL}
\end{figure}

%%%%%%%%%%%%%%%%%%
\medskip
\noindent \underline{2HDM-II with SMF + VLL}:
To the Type-II 2HDM we add one doublet VLL $\psi_l$ with hypercharge $Y_{\psi_l}$, and one singlet VLL $\chi$ with hypercharge $(Y_{\psi_l}-1/2)$.
We couple the VLLs to $\Phi_1$ in the same way as in the $MVLE_{11}$~model of Ref.~\cite{Gopalakrishna:2015wwa}, with $\tilde{y}$~s set equal to zero.
The coupling of $\Phi_1$ to the VLLs will be denoted by $y^l_1$.
After EWSB the $\chi$ and the
lower component of $\psi_l$ mix to produce two mass eigenstates which we call $\zeta_1$ and $\zeta_2$ in accordance with Ref.~\cite{Gopalakrishna:2015wwa}
where $\zeta_2$ is the lighter eigenstate.
The effective $\phi ff$ couplings, i.e $y^\phi_{ij}$s in notation of Ref.~\cite{Gopalakrishna:2015wwa}, 
and the mass eigenvalues can be found in App.~A of Ref.~\cite{Gopalakrishna:2015wwa}.
We take  $Y_{\psi_l}=-1/2$ and choose the mass parameters of the VLLs such that
the lighter mass eigenvalue of the charge $-1$ VLL is $375~$GeV.
In Fig.~\ref{sigmabr-2HDM.VLL} we plot $\sigma_\phi * BR_{\gamma \gamma}$ as a function of $y^l_1$
for various values of $\{ \tan \beta, \kappa^2_{\Gamma}\}$, 
and also the unitarity constraint from $\psi \psi \to \psi \psi$ process given by $\sqrt{2}(y^H_{22} + y^A_{22}) < 10$ as a red vertical line.
We can see that within the unitarity constraint the maximum $\sigma*BR_{\gamma \gamma} \simeq 0.5~$fb for $\kappa^2_{\Gamma} = 0.24$.
%

%%%%%%%%%%%%%%
\medskip
\noindent \underline{2HDM-II with SMF + VLQ + VLL}:
To the Type-II 2HDM we introduce one doublet VLQ $\psi_q$ with hypercharge $Y_{\psi_q}$, one singlet VLQ $\xi$ with hypercharge $Y_{\psi_q} + 1/2$,
one doublet VLL $\psi_l$ with hypercharge $Y_{\psi_l}$,  and one singlet VLL $\chi$ with hypercharge $(Y_{\psi_l}-1/2)$.
We couple the VLQs and the VLLs to the scalar doublets in the same way as in the $MVQU_{22}$~model and the $MVQD_{11}$~model of Ref.~\cite{Gopalakrishna:2015wwa} respectively,
with $\tilde{y}$~s set equal to zero. 
The couplings of the scalar doublets with the VLQs and the VLLs will be denoted by $y^q_1$ and $y^l_1$ respectively.  
The effective $\phi ff$ couplings and the mass eigenvalues can be found in App.~A of Ref.~\cite{Gopalakrishna:2015wwa}.
We take $N'_c =3$, $Y_{\psi_q}=1/6$, $Y_{\psi_l}=-1/2$ and choose the mass parameters of the VLFs such that
the lighter mass eigenvalues of the charge $2/3$ VLQs and the charge $-1$ VLLs is $1000~$GeV and $375~$GeV respectively.
%%%%%%%%%%%%%%%%%%%%%%%%%%%%%%%%%%%%%%%%%%%%%%%%%%

\begin{figure}
\centering
\includegraphics[width=0.32\textwidth]{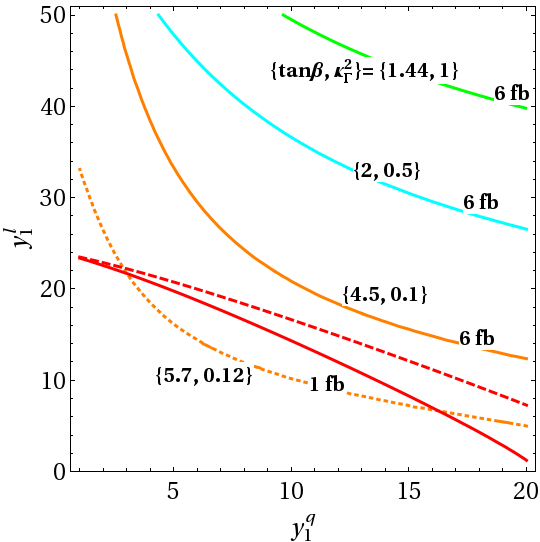}
\includegraphics[width=0.32\textwidth]{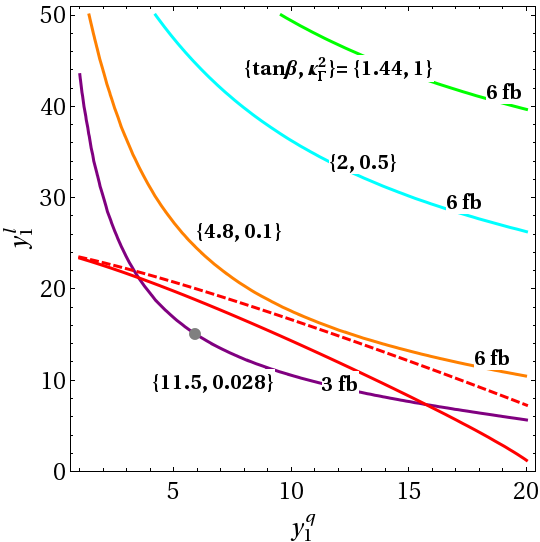}
\includegraphics[width=0.32\textwidth]{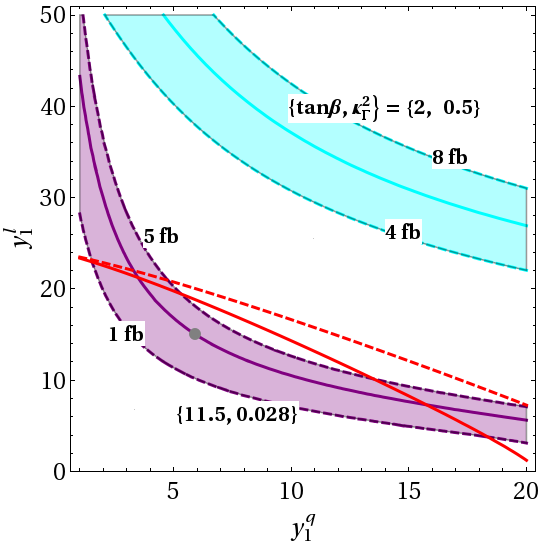}
\caption{In 2HDM-II (left) and 2HDM-X (middle and right) with only $H$ present,
  and VLF mass parameters chosen such that the lighter mass eigenvalues of the VLQs and the VLLs is $1000~$GeV and $375~$GeV respectively. 
The unitarity constraint from $\phi \phi \to \phi \phi$ and $\psi \psi \to \psi \psi$ is shown in solid red and dashed red respectively.
The upper limit on $y^q_1$ from 8~TeV LHC $\phi \to \tau \tau$ result is shown by gray dots.
}
\label{yQpyLp}
\end{figure}

For illustration, we start by including only the $H$ contribution, 
and show in Fig.~\ref{yQpyLp} the values of $y^l_1, y^q_1$ needed to explain the $750~$GeV excess for various $\{ \tan \beta ,\kappa_{\Gamma}^2 \}$.
We also show in Fig.~\ref{yQpyLp}, the unitarity constraint on $y^q_1, y^l_1$
from $\phi \phi \to \phi \phi$ (shown in solid red) and $\psi \psi \to \psi \psi$ (dashed red) channels
given by the equations  $[y^q_{22} (N'_c)^{1/4} + y^l_{22}] < 10$ and $(y^q_{22} + y^l_{22}) < 10$ respectively,
where $y^q_{22}$ and $y^l_{22}$ are the couplings of the $H$ to the lighter VLQ and the lighter VLL respectively.
In this case the 8~TeV LHC $\phi \to tt$ results do not put any additional constraints.
We see that if the unitarity constraint is to be satisfied, it is not possible to explain the excess in 2HDM-II with only $H$ contributing to $\sigma*BR$.
Within the unitarity bound, the maximum $\sigma*BR_{\gamma \gamma} = 1$~fb for $\kappa^2_{\Gamma} =0.12$.

Next we include both $H$ and $A$ contributions to $\sigma*BR_{\gamma \gamma}$, and show
in Fig.~\ref{sigmabr-2HDM.VLLQ} the values of $y^q_1,y^l_1$ required to explain the excess
by showing the region $4 < \sigma_\phi * BR_{\gamma \gamma} < 10$~fb for various $\{ \tan \beta, \kappa^2_{\Gamma} \}$.
We also show in Fig.~\ref{sigmabr-2HDM.VLLQ} the unitarity constraint on $y^l_1, y^q_1$ from
$\phi \phi \to \phi \phi$ (solid red), $\psi \psi \to \psi \psi$ (dashed red) processes.
The unitarity constraint from $\psi \psi \to \psi \psi$ is now
$2^{1/4}[y^q_{22} + y^l_{22} ] < 10$. 
Since both $BR_{tt}$ and $\kappa^2_\Gamma$ are largely controlled by $\tan \beta$, a given $\kappa^2_\Gamma$ implies a certain $BR_{tt}$, and in 
fig.~\ref{sigmabr-2HDM.VLLQ} we show by thick red dots the upper limit on $y^q_1$ for a given $\kappa^2_\Gamma$ from the LHC $t\bar t$ search limits.
For $\kappa^2_\Gamma \gtrsim 0.5$ ($\tan \beta < 3$), BR$(\phi \to tt) > 0.9$;
therefore we get a nontrivial constraint on $y^q_1$ in this region.
For $\kappa^2_\Gamma \simeq 0.24$ ($\tan \beta \simeq 5.7$), the $BR(\phi \to tt)$ is reduced to $ \simeq 0.5$ and
we do not get any constraint on $y^q_1$ from the $\phi \to t t$ results in the range we consider.
We see that in this case it is possible to generate $\sigma*BR_{\gamma \gamma} = 6$~fb for $\kappa^2_\Gamma \simeq 0.24$,
without violating the unitarity constraint and the constraints from 8~TeV LHC $\phi \to tt$ searches.
For $\sigma*BR_{\gamma \gamma} =6$ fb the maximum $\kappa^2_\Gamma \simeq 0.5$.
The reason for the larger cross section is the inclusion of $A$ contribution;
$\kappa_{A VV} \approx 2.5*\kappa_{H VV}$ when $M_f \simeq M_\phi/2$ with $M_f$ the mass of the fermion in the loop, 
and about $1.33* \kappa_{HVV}$ for $M_\psi \gg M_\phi$ (see for example Ref.~\cite{Djouadi:2005gj}).

%%%%%%%%%%%%%%%%%%%%%%%%%%%
\subsubsection{2HDM type-X}
In the 2HDM type-X model (see Ref.~\cite{Chang:2012ve} for a review), which we call 2HDM-X, 
the SM quarks couple to $\Phi_2$ and their couplings to $H,A$ are proportional to $m_f \cot \beta$,
while the SM leptons couple to $\Phi_1$ and their couplings to $H,A$ are proportional to $m_f \tan \beta$.

We find that the minimum $\kappa^2_{\Gamma}$ is $\simeq 0.028$ which occurs for $\tan \beta =11.5$, 
which is smaller than 2HDM-II because of the smallness of $M_\tau$ compared to $M_b$.
We add the VLQs and the VLLs in the same way as we did for the 2HDM-II model.
For illustration, we include only the $H$ contribution to the diphoton process,
and show in Fig.~\ref{yQpyLp} the parameter values needed to explain the excess for various $\kappa_{\Gamma}^2$,
the region $4< \sigma_H*BR_{\gamma \gamma}<8$~fb around $\{ \tan \beta, \kappa^2_\Gamma \} = \{ 2, 0.5 \}$
and the region $1< \sigma_H*BR_{\gamma \gamma}<5$~fb around  $\{ \tan \beta, \kappa^2_\Gamma \} = \{ 11.5, 0.028 \}$
with the VLF parameters taken the same as in 2HDM-II.
The unitarity bound from $\phi \phi \to \phi \phi$ and $\psi \psi \to \psi \psi$ are also shown in Fig.~\ref{yQpyLp}
by the solid red and dashed red curves respectively.
Since in 2HDM-X $BR(\phi \to \tau \tau)$ can become large for large $\tan \beta$, the 8~TeV LHC constraints from $\phi \to \tau \tau$ channel gives additional constraints on $y^q_1$.
For $\tan \beta =11.5$, $BR(H \to \tau \tau) \simeq 0.46$, which gives an upper bound on $y^q_1 \simeq 5.9$ as shown in Fig.~\ref{yQpyLp} by the gray dot.
We see that within the unitarity bound and 8~TeV LHC $\phi \to \tau \tau$ constraint $\sigma*BR_{\gamma \gamma} \simeq 1$~fb 
can be obtained for $\kappa^2_{\Gamma} =0.028$.

Next we include both $H$ and $A$ contributions to $\sigma*BR_{\gamma \gamma}$, and 
show in Fig.~\ref{sigmabr-2HDM.VLLQ} the values of $y^q_1,y^l_1$ required to explain the excess
and the region $4< \sigma*BR_{\gamma \gamma}<10$~fb for various $\{ \tan \beta, \kappa^2_{\Gamma} \}$.
An upper limit on $y^q_1$ from the 8~TeV LHC $\phi \to tt$ results for a given $\kappa^2_\Gamma$ are also shown in Fig.~\ref{sigmabr-2HDM.VLLQ} by the thick red dots.
In this case we get an upper limit of $y^q_1 \simeq 13.5$ for $\tan \beta =6.8$, $\kappa^2_\Gamma = 0.1$,
while in 2HDM-II we did not get any bound for the nearby value of $\tan \beta =5.7$, $\kappa^2_{\Gamma} =0.24$.
The difference between these two cases comes from the fact that in 2HDM-X,
the $\phi bb$ coupling is also suppressed by $1/\tan \beta$
so that $BR(\phi \to tt) \simeq 0.9$ even for $\tan \beta = 6.8$.
The upper limit on $y^q_1$ from the 8~TeV LHC $\phi \to \tau \tau$ result is also shown in Fig.~\ref{sigmabr-2HDM.VLLQ} by the gray dot.
For $\tan \beta =6.8$, $BR(\phi \to \tau \tau) \simeq 0.1$ and the upper limit is $y^q_1 \simeq 6.5$ in this case.
We see that in this case it is possible to generate $\sigma*BR_{\gamma \gamma} =6$~fb for $\kappa^2_\Gamma \simeq 0.1$.
For $\sigma*BR_\gamma = 6$~fb a maximum of $\kappa^2_\Gamma \simeq 0.5$ can be reached in this model as in the 2HDM-II.
%
%%%%%%%%%%%%%%%%%%%%%%%%%%%%%%%%%%%%%%%%%%%%%%%%%%%%%%%%%%%%%%%%%%%%%%%%%%%%%%%%%%%%%%%%

%%%%%%%%%%%%%%%%%%%%%%%%%%
\subsubsection{2HDM type-I}
In the 2HDM type-I model (which we call 2HDM-I) all the SM fermions couple to $\Phi_2$, 
and hence all the SM fermions couple to $H,A$ proportional to $\cot\beta$.
In this case the 8~TeV LHC $\phi \to \tau \tau$ limits do not put any constraints on the parameter space
and $\Gamma_\phi$ can be made very low by going to large $\tan \beta$.
We expect that the addition of VLFs increases $\sigma*BR_{\gamma \gamma}$. 
We first consider the case when only VLLs are added, and subsequent to this when both VLLs and VLQs are added.

%%%%%%%%%%%%%%%%%
\medskip
\noindent \underline{2HDM-I with SMF + VLL}:
We introduce VLLs in the same way as we did in the 2HDM-II + VLL model.
We again take $Y_{\psi_l}=-1/2$ and choose the mass parameters of the VLLs such that
the lighter mass eigenvalue of the charge $-1$ VLL is $375~$GeV.
Including both $H$ and $A$ contributions to $\sigma*BR_{\gamma \gamma}$, we 
show in Fig.~\ref{sigmabr-2HDM.VLL} the $\sigma*BR_{\gamma \gamma}$ as a function of $y^l_1$
for various values of $\{ \tan \beta, \kappa^2_{\Gamma}\}$.
We also show in Fig.~\ref{sigmabr-2HDM.VLL} the unitarity constraint on $y^l_1$ from the $\psi \psi \to \psi \psi$ process as a red vertical line.
We see that in this case, $\sigma*BR_{\gamma \gamma} \gtrsim 10$~fb can be comfortably reached within the unitarity constraint, albeit for small $\kappa^2_{\Gamma}$.

\begin{figure}
\centering
\includegraphics[width=0.4\textwidth]{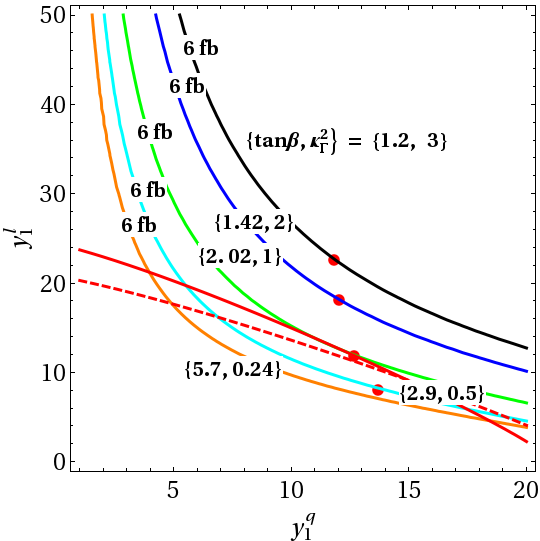}
\includegraphics[width=0.4\textwidth]{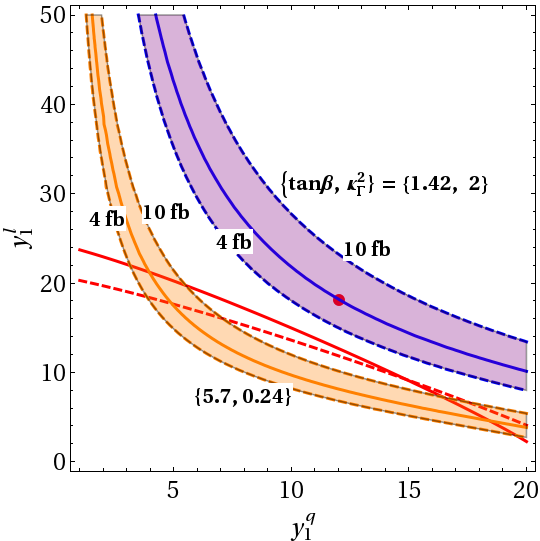} \\
\includegraphics[width=0.4\textwidth]{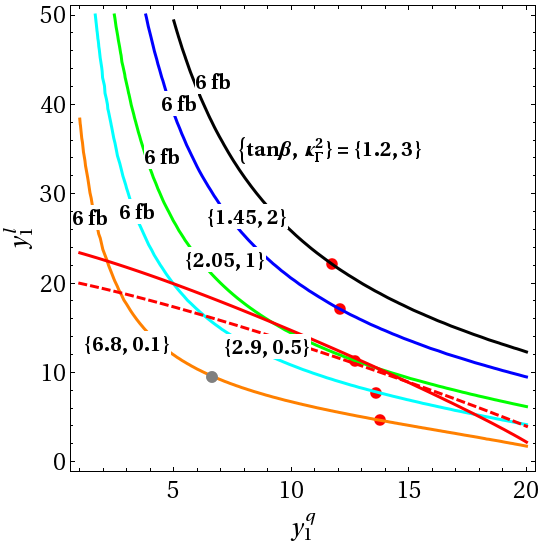}
\includegraphics[width=0.4\textwidth]{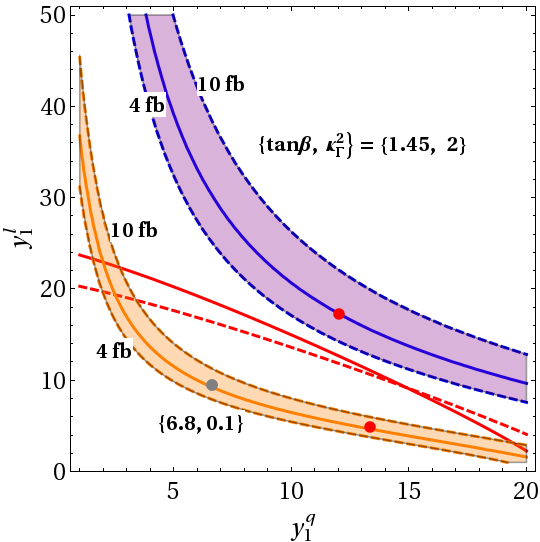} \\
\includegraphics[width=0.4\textwidth]{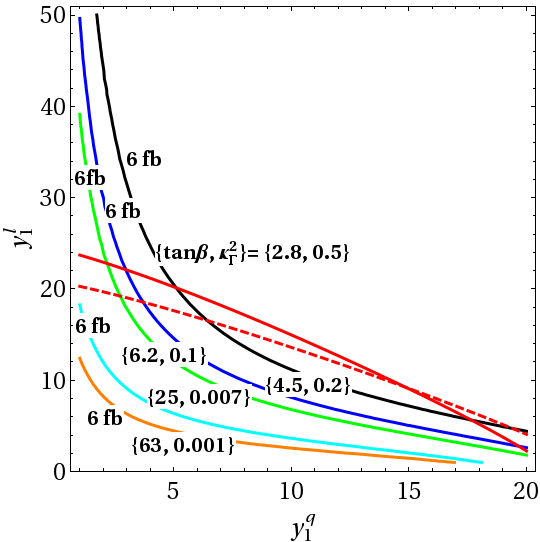}
\includegraphics[width=0.4\textwidth]{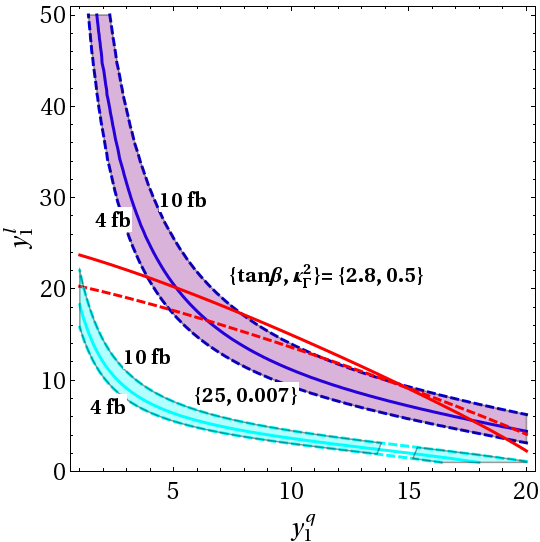}
\caption{In 2HDM-II (top panel), 2HDM-X (middle panel) and 2HDM-I (bottom-panel) for SMF + VLL + VLQ,
with $M_H, M_A = 735, 750$~GeV and VLF mass parameters chosen such that the lighter mass eigenvalues of the VLQs and the VLLs is $1000~$GeV and $375~$GeV respectively,
unitarity constraint from $\phi \phi \to \phi \phi$ (solid red), $\psi \psi \to \psi \psi$ (dashed red).
To the right of the red dots and the gray dots are excluded from 8~TeV LHC $\phi \to tt$ data and the 8~TeV LHC $\phi \to \tau \tau$ data respectively.
}
\label{sigmabr-2HDM.VLLQ}
\end{figure}

%%%%%%%%%%%%%%%%%
\medskip
\noindent \underline{2HDM-I with SMF + VLL +VLQ}:
In addition to the SMF in 2HDM-I, we add VLL and VLQ in the same way as we did in 2HDM-II + VLL + VLQ model.
As before we take $Y_{\psi_q}=1/6$, $Y_{\psi_l}=-1/2$ and choose the mass parameters of the VLFs such that the lighter mass eigenvalues of the charge $2/3$ VLQs and the charge $-1$ VLLs
is $1000~$GeV and $375~$GeV respectively.
In Fig.~\ref{sigmabr-2HDM.VLLQ} we show contours of $\sigma*BR_{\gamma \gamma}$
and the region $4<\sigma*BR_{\gamma \gamma} < 10 $~fb with $M_H = 735~$GeV, $M_A =750~$GeV,
for various values of $\{ \tan \beta, \kappa^2_{\Gamma}\}$.
We also show in Fig.~\ref{sigmabr-2HDM.VLLQ} the unitarity constraint on $y^l_1, y^q_1$
from $\phi \phi \to \phi \phi$ and $\psi \psi \to \psi \psi$ processes (shown here by solid red and dashed red respectively).

We see that in this model $\sigma*BR_{\gamma \gamma} \simeq 10$~fb can be reached within the unitarity constraints for $\kappa^2_\Gamma \simeq 0.5$.

%%%%%%%%%%%%%%%%%%%%%%%%%%%%%%%%%%%%%%%%%%%%%%
\subsection{Electroweak singlet $\phi$}
\label{singPhiHidDM.SEC}

We explore here the possibility of the 750~GeV resonance being an SU(2) singlet scalar $\phi$.
The large width of the $\phi$ can be due to $\phi\to \psi\psi$ decays, where $\psi$ is a vector-like BSM fermion, which if EM neutral could be a dark matter candidate.  
We take the $\phi$ to be CP-even in this work.
A coupling between the vector-like fermionic dark matter $\psi$ and the SM sector can arise via the Higgs-portal due to a mixing between the $\phi$ and the SM Higgs boson.
This mixing is possible only for a CP-even scalar if $CP$-invariance is not to be broken spontaneously.
Our diphoton channel results, although presented for a CP-even scalar, apply qualitatively also to a CP-odd scalar,
but the exact values of the CP-odd scalar couplings preferred will be different due to ${\cal O}(1)$ factor differences in the
$\phi gg$ and $\phi \gamma\gamma$ loop factors for the CP-even and CP-odd scalar cases. 

We introduce one SU(2) singlet scalar $\hat\phi$,
with an SU(2) singlet color triplet VLQ $U$ with hypercharge $Y_U$ and mass $M_U$,
and an $SU(2)$ singlet VLF $\psi$ with mass $M_\psi$.
This model and the couplings to the VLF parallels the SVU model of Ref.~\cite{Gopalakrishna:2015wwa},
and in the notation of that paper we refer to this model as the $SVU\psi$ model. 
Without committing ourselves to a particular theory, we write an effective theory 
\beq
\Lagr \supset - M_h^2 H^\dagger H - M_\phi^2 \Phi^\dagger \Phi 
- \kappa \Phi^\dagger \Phi H^\dagger H - \mu \Phi H^\dagger H
- M_\psi \bar\psi \psi - M_U \bar{U} U - \frac{y_\psi}{\sqrt{2}} \hat\phi \bar\psi \psi - \frac{y_U}{\sqrt{2}} \hat\phi \bar U U 
\ ,
\label{SVUpsiLagr}
\eeq
We assume that the potential is such that $\left< \Phi \right> = \xi/\sqrt{2}$ and $\left< H \right> = v/\sqrt{2}$,
and denote the fluctuations around these as $\hat \phi$ and $\hat h$ respectively.
The effective coupling $\kappa_{\phi h h}$ defined in Eq.~(\ref{phihhDefn.EQ}) is given as $\kappa_{\phi h h} = \sqrt{2} (\mu + \kappa\xi)/M_\phi$. 
The $\hat\phi$ and $\hat h$ mix after EWSB and the mixing angle $\sin\theta_h \equiv s_h$ is given by
\beq
\tan(2\theta_h) = \frac{\sqrt{2} \kappa_{\phi hh} }{\left(1 - M_h^2/M_\phi^2\right)} \frac{v}{M_\phi}  \ ,
\label{t2th.EQ}
\eeq
with the effective coupling $\kappa_{\phi hh}$ defined in Eq.~(\ref{SVUpsiLagr}). 
Diagonalizing the $\hat\phi \leftrightarrow \hat{h}$ mixing terms, we go from the $(\hat h, \hat\phi)$ basis to the mass basis $(h,\phi)$, 
and define the mass eigenstates to be $h = c_h \hat h - s_h \hat \phi$ and $\phi = s_h \hat h + c_h \hat \phi$. 
In the $(\phi, h)$ mass basis we have  
\beq
{\cal L}_{\phi h h} = -\frac{1}{4} \tan{2\theta_h} (c_h^3 - 2 c_h s_h^2) \frac{(M_\phi^2 - M_h^2)}{v} \phi h h \ .
\eeq
In our numerical analysis below, we treat $s_h$ as an input parameter, and one can always relate it to the $\Lagr$ parameters in a model if needed using Eq.~(\ref{t2th.EQ}).
In order to agree with the Higgs observables already measured at the LHC, $s_h$ must be small as we show later. 
The phenomenology of the $\kappa$ is discussed in detail for example in Ref.~\cite{Gopalakrishna:2009yz}.
For example, for $s_h = 0.01$, we have $\kappa_{\phi h h} = 0.04$. 

Mixed operators such as
\beq
\Lagr_{\rm mix} \supset -\tilde{y}_U \bar U q^3_L \cdot H -\tilde{y}_\psi \bar \psi \ell^3_L \cdot H + {\rm h.c.} \ ,
\label{LZ2br.EQ}
\eeq
are allowed if $Y_U = 2/3$ and $Y_\psi = 0$, where $q^3_L$ is the third-generation SM quark doublet, and $\ell^3_L$ is the third-generation SM lepton doublet.
To be safe from FCNC constraints, we allow couplings with only third-generation SM fermions. 
To prevent having a cosmologically stable $U$, we take $\tilde{y}_U$ to be small enough that all FCNC constraints are obeyed,
but big enough that $U$ decays promptly to SM final states as discussed in detail in Ref.~\cite{Gopalakrishna:2006kr}, and we do not therefore discuss further the consequences of this operator in this work.
The $\Lagr$ respects a $Z_2$ symmetry under which $\psi \to -\psi$, and this $Z_2$ symmetry is broken only by the $\tilde{y}_\psi$ term.
Thus, if $\tilde{y}_\psi = 0$, the $\psi$ is absolutely stable and is a possible dark matter candidate.
One then has to ensure that the parameters are chosen in such a way that the relic density is not so high that it over-closes the universe, or the direct-detection cross-section is not so high that it is excluded by experiment. 
We explore this possibility in detail below. 

\begin{figure}
\centering
\includegraphics[width=0.32\textwidth]{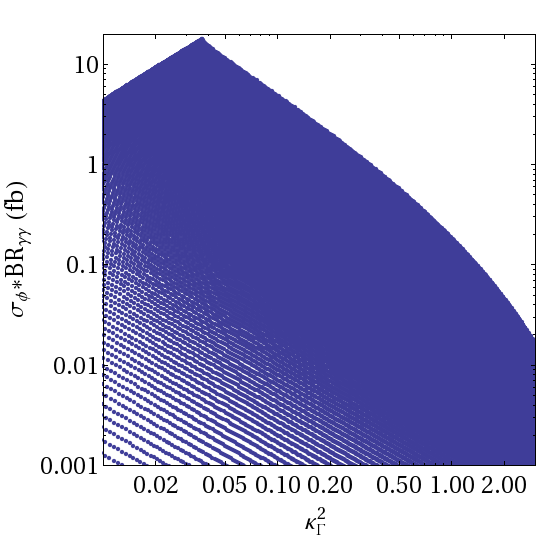}
\caption{In the $SVU\psi$ model for $Y_\psi = 0$, $Y_U= 2/3$, the $\sigma_\phi*BR_{\gamma \gamma}$ vs. $\kappa_\Gamma^2 $, for $M_\psi =350~$GeV, $M_U= 1000~$GeV, $s_h=0.01$
  and $y_U, y_\psi$ scanned over the range  $0 < y_U < y_U^{max}$, $0 < y_\psi < y_\psi^{max}$  subject to the unitarity constraint.
}
\label{sigmabrScan.svupsi}
\end{figure}

The $\sigma_\phi*BR_{\gamma \gamma}$ can be obtained from Eq.~(\ref{sigmaphiBrGm}),
and the expressions for $\kappa_{\phi gg }$ and $\kappa_{\phi \gamma\gamma}$ are given in App.~B of Ref.~\cite{Gopalakrishna:2015wwa}.
In $\Gamma_\phi$ we include the partial widths $\Gamma(\phi \to \psi \psi, hh,tt,gg)$.
In Fig.~\ref{sigmabrScan.svupsi} we show $\sigma_\phi*BR_{\gamma \gamma}$ vs. $\kappa_\Gamma^2$ in the $SVU\psi$ model,
for $M_\psi =350~$GeV, $M_U= 1000~$GeV, $Y_U = 2/3$, $Y_\psi = 0$, $s_h = 0.01$ and scanning over $y_U,y_\psi$
in the range $0 < y_U < y_U^{max}$, $0 < y_\psi < y_\psi^{max}$, subject to the unitarity constraint $y_\psi + y_U N_c^{1/4} < 10$ computed in Sec.~\ref{utrtyCon.SEC}.
The 8~TeV $hh$ channel constraints discussed in Sec.~\ref{constr.SEC} constrains $\kappa_{\phi hh} \ll 1$. 
For instance this implies the bound $s_h \lesssim 0.05$ for $y_U = 5$ and $\kappa_\Gamma^2 = 0.1$. 
For $y_\psi \gtrsim 0.1 $, the $BR(\phi \to \psi\psi)$ is dominant and $y_\psi$ largely controls $\kappa_\Gamma^2$. 
For $\kappa_\Gamma^2 = 3$, the $\sigma_\phi \times BR_{\gamma\gamma}$ can reach only $0.01$~fb for $s_h = 0.01$ as seen in the left plot.
For very small $y_\psi \lesssim 0.1$, the total width (i.e. $\kappa_\Gamma^2$) is small and dominated by top and $U$ loops and the tree-level $\phi \to hh,tt$ decays.
For $y_\psi \to 0$, $s_h \to 0$ both $\sigma*BR_{\gamma \gamma}$ and $\kappa^2_\Gamma$ comes from $U$ loops and scales as $y_U^4$ and $y_U^2$ respectively;
$\sigma*BR_{\gamma \gamma}$ increases with $\kappa^2_\Gamma$ in this region up to around $\kappa^2_\Gamma \simeq 0.03$ as can be seen from Fig.~\ref{sigmabrScan.svupsi}.
We can see that for $s_h=0.01$ we can get $\sigma*BR_{\gamma \gamma} \simeq 10$~fb.

\begin{figure}
\centering
\includegraphics[width=0.32\textwidth]{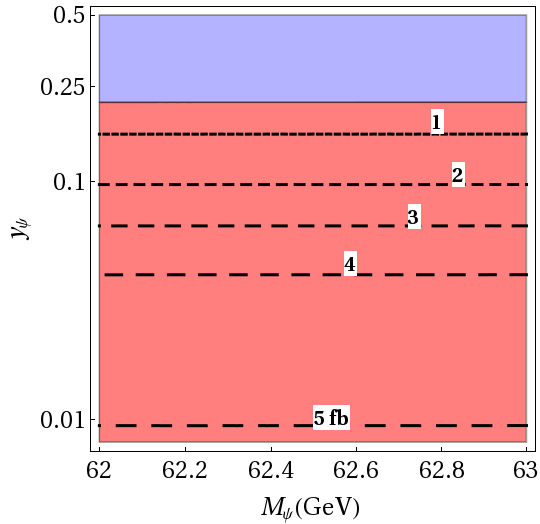}
\includegraphics[width=0.32\textwidth]{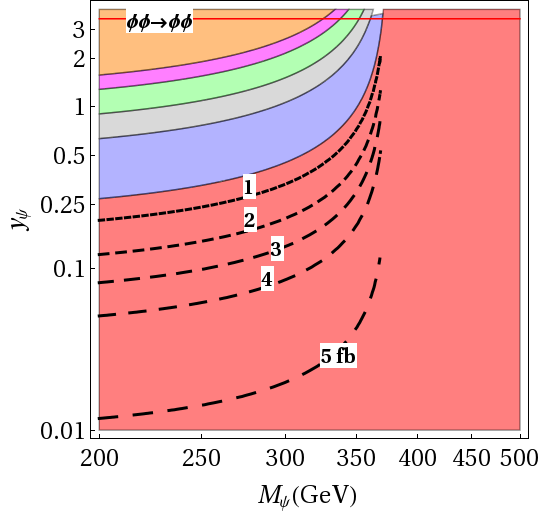}\\
\includegraphics[width=0.32\textwidth]{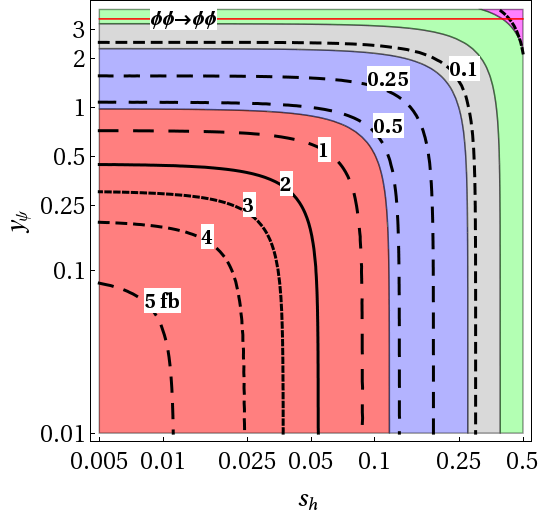}
\includegraphics[width=0.32\textwidth]{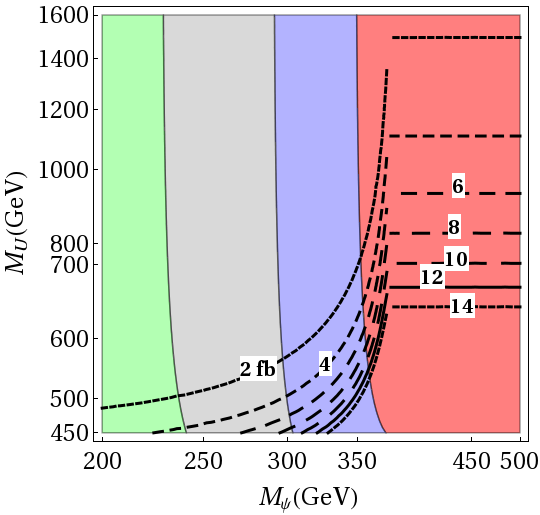}
\caption{In the $SVU\psi$ model for $Y_\psi = 0$, $Y_U = 2/3$,
the contours of $\sigma_\phi*BR_{\gamma \gamma} $ (in fb), 
and regions of
$\kappa_\Gamma^2 < 0.1$ (red),
$0.1<\kappa_\Gamma^2 < 0.5$ (blue),
$0.5<\kappa_\Gamma^2 < 1$ (gray),
$1<\kappa_\Gamma^2 < 2$ (green),
$2<\kappa_\Gamma^2 < 3$ (pink),
$\kappa_\Gamma^2 > 3$ (orange);  
parameters not along the axes are fixed at $s_h =0.01$, 
$M_\psi = 350$~GeV, $M_U = 1000$~GeV, 
$y_\psi = 1$, $y_U=5$.
Unitarity constraint on $y_\psi$ for $y_U=5$ is shown by the red horizontal line.
}
\label{sigmabr.svupsi}
\end{figure}

In Fig.~\ref{sigmabr.svupsi} we show contours of
$\sigma_\phi*BR_{\gamma \gamma} $ (in fb), 
and various $\kappa_\Gamma^2$ as colored regions, with the 
parameters not along the axes fixed at $s_h =0.01$, 
$M_\psi = 350$~GeV, $M_U = 1000$~GeV, 
$y_\psi = 1$, $y_U=5$.
We also show in Fig.~\ref{sigmabr.svupsi} the unitarity constraint on $y_\psi$ (shown here by red line) for $y_U = 5$.
For $y_U =5$, $\sigma_\phi \simeq 1.5$~pb and
The partial widths $\Gamma_{\{hh,tt,gg\}}$ for $s_h =0.01$, $M_U = 1000~$GeV are
$0.0065,0.0031,0.16$~GeV respectively.
For very small $y_\psi$ or $M_\psi > M_\phi/2$, $\Gamma(\phi \to \psi \psi) \simeq 0$
and $\Gamma_\phi$ is dominated by $\Gamma_{\{hh,tt,gg\}}$;
in this limit $BR_{\gamma \gamma} \simeq 3.3*10^{-3}$ and $\sigma*BR_{\gamma} \simeq 5$~fb
for the set of parameters chosen with $s_h = 0.01$.
If we decrease $M_U$, $\sigma*BR_{\gamma \gamma}$ can be even larger;
for $M_U \simeq 650$~GeV, $M_\psi > M_\phi/2$,  $\sigma*BR_{\gamma \gamma} \simeq 12$~fb can be reached as can be seen from Fig.~\ref{sigmabr.svupsi}
although for a small $\kappa^2_\Gamma \simeq 0.03$.
For $M_\psi < M_\phi/2$ and $y_\psi$ large, $\Gamma_\phi$ is large being dominated by $\phi \to \psi \psi$ decay
resulting in very small $\sigma*BR_{\gamma \gamma}$.
Thus, in the $SVU\psi$ model, it is not possible to generate {\em both} a large $\sigma_\phi \times BR_{\gamma\gamma}$ of a few fb and also a large $\kappa_\Gamma^2 \approx 3$.
The reason is simply because a large $\Gamma$ corresponding to $\kappa_\Gamma^2 \approx 3$ suppresses the $BR_{\gamma\gamma}$ to tiny values.

\begin{figure}
\centering
\includegraphics[width=0.32\textwidth]{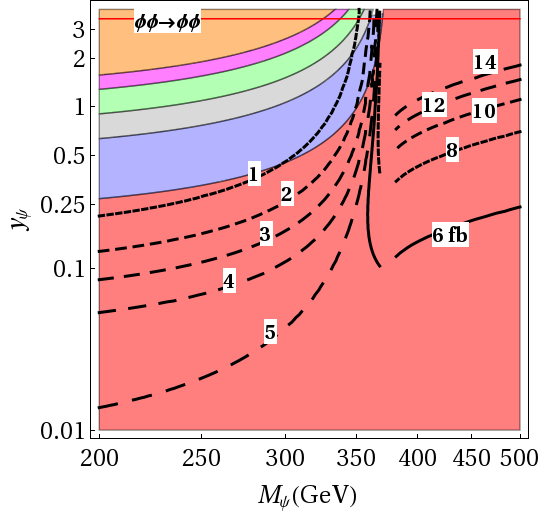}
\includegraphics[width=0.32\textwidth]{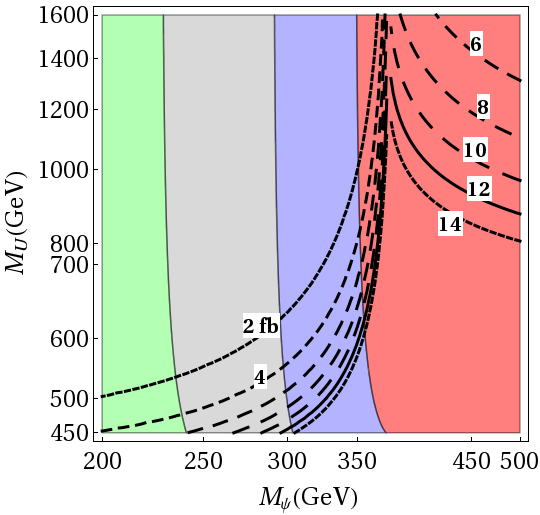}
\caption{In the $SVU\psi$ model for $Y_\psi = -1$, $Y_U =2/3$, $s_h=0$, $y_U=5$, 
contours of $\sigma*BR_{\gamma \gamma} =  0.1,~0.5,~1,~2,~4,~6,~8$ fb, showing the regions
 $\kappa_\Gamma^2 < 0.1$ (red),
 $0.1<\kappa_\Gamma^2 < 0.5$ (blue),
 $0.5<\kappa_\Gamma^2 < 1$ (gray),
 $1<\kappa_\Gamma^2 < 2$ (green),
 $2<\kappa_\Gamma^2 < 3$ (pink),
 $\kappa_\Gamma^2 > 3$ (orange), 
with $M_U = 1000~$GeV (left),
and $y_\psi = 1$ (right).
Unitarity constraint on $y_\psi$ from $\phi \phi \to \phi \phi$ for $y_U=5$ is shown by the red horizontal line.
}
\label{sigmabr.svul}
\end{figure}

%%%%%%%%%%%%%%%%%%%%%%%
We could take $Y_\psi = -1$, and since $\psi$ is an SU(2) singlet, it has EM charge $Q_\psi = Y_\psi = -1$. 
For this case, to prevent a cosmologically stable charged relic, we additionally include a mixing term to a SM lepton that allows $\psi$ to decay.
Of course in this case $\psi$ cannot be dark matter.
One such example of a mixing term is to the SM SU(2) singlet $\tau_R$, namely, $\Lagr \supset - M_{\psi\tau}' \psi \tau^c + {\rm h.c.}$, 
with $M_{\psi\tau}'$ taken small enough that leptonic FCNC constraints are not violated,
but large enough that the $\psi$ decay life-time due to $\psi \to h \tau$ decays is much smaller than cosmological time scales. 
Since $\psi$ has EM charge, it will contribute to $\Gamma_{\gamma \gamma}$ also.
In Fig.~\ref{sigmabr.svul} we show for $Y_\psi=-1$, contours of $\sigma_\phi*BR_{\gamma \gamma}$
and regions of $\kappa_\Gamma^2$
for parameters not shown along the axes fixed at $y_\psi = 1$, $y_U=5$, $s_h=0$, $M_U = 1000$~GeV.
We also show in Fig.~\ref{sigmabr.svul} the unitarity constraint on $y_\psi$ from $\phi \phi \to \phi \phi$
process for $y_U =5$.
As explained earlier, for $s_h = 0$, in the region $M_\psi > M_\phi/2$, $\Gamma_\phi \approx \Gamma_{gg}$ is small 
and therefore $BR_{\gamma \gamma}$ can be sizable, and $\sigma*BR_{\gamma \gamma} \approx 8~$fb is reached, albeit for $\kappa_\Gamma^2 \ll 0.1$.
For $M_\psi \approx M_\phi/2$, the $\kappa_{\phi\gamma\gamma}$ loop function is enhanced as seen in Fig.~\ref{sigmabr.svul}.  

%%%%%%%%%%%%%%%%%%%
\subsubsection{Hidden sector dark matter}
\label{HidSec.SEC}
If $\tilde{y}_\psi$ of Eq.~(\ref{LZ2br.EQ}) is zero, the $Z_2$ symmetry is exact, $\psi$ is stable and can potentially be a dark matter candidate for $Y_\psi = 0$.
The dark matter relic density and direct-detection
can be computed as detailed, for example, in Ref.~\cite{Gopalakrishna:2009yz} and App.~A of Ref.~\cite{Gopalakrishna:2006kr}.
In order to get the correct relic density of $\Omega_{dm} = 0.26 \pm 0.015$~\cite{Adam:2015rua},
we need the thermally averaged self-annihilation cross-section to be $\left< \sigma v\right> \approx 2.3\times 10^{-9}~{\rm GeV}^{-2}$.
We have for our case~\cite{Gopalakrishna:2009yz,Gopalakrishna:2006kr}
\beq
\left< \sigma v\right> = \frac{6}{x_f} \frac{1}{8\pi s} \sum_i  |\mathcal{B}_i|^2  \hat\Pi_{PS}^i \ ,
\eeq
where $x_f \equiv M_\psi/T_f \approx 25$ with $T_f$ the freeze-out temperature, the sum is over all self-annihilation processes $\psi\psi \to f_i f_i$ for final states $f_i$ kinematically allowed,
the $|\mathcal{B}_i|^2$ is the coefficient of $v_{rel}^2$ in the amplitude squared for each process,
$v_{rel}$ being the relative velocity of the two initial state $\psi$;
the $\hat\Pi_{PS}^i \equiv \sqrt{(1-4m_i^2/s)}$ is a phase-space factor with $m_i$ the mass of the final-state particle, and $s$ is the Mandelstam variable, which
for a cold-dark matter candidate during freeze-out is $s\approx 4 M_\psi^2$.
In our analysis we include the two-body final states $b\bar b, WW, ZZ, hh, t\bar t, gg$, whichever are kinematically allowed for that given $M_\psi$.
Although $\tau \tau$ and $\gamma \gamma$ final states are also possible,
we ignore them in our analysis as these contributions are small.
For large $s_h$, the loop level $gg$ contribution is small compared to other tree level contribution.
But for small $s_h$, $gg$ contribution becomes comparable or even larger than the tree level processes.
Details of the $\mathcal{B}_i$ for each of these final states are given in Appendix~\ref{hidRelDen.APP}.  

The dark-matter direct-detection elastic scattering cross-section on a nucleon is
mediated by scalar exchange.
Since $h$ is lighter than $\phi$, the former mostly contributes, but if $s_h \lesssim 0.05$, 
the latter's contribution is also important. 
The $h$ exchange contribution is given for example in Ref.~\cite{Gopalakrishna:2009yz}, which 
we generalize here to include $\phi$ contribution also since we consider $s_h \lesssim 0.05$. 
The scalar-nucleon-nucleon coupling is generated due to the scalar coupling to the quark content of the nucleon, 
and also due the scalar coupling to the gluon content of the nucleon via the $ggh,gg\phi$ effective couplings.  
We define an effective Lagrangian for the scalar-nucleon-nucleon interaction as
\bea
\Lagr &\supset& \lambda_{hNN} \hat{h} \bar{N} N + \lambda_{\phi NN} \hat{\phi} \bar{N} N \ , \nonumber \\
         &=& (c_h \lambda_{hNN} - s_h \lambda_{\phi NN}) h \bar{N} N + (c_h \lambda_{\phi NN} + s_h \lambda_{hNN} ) \phi \bar{N} N \ , 
\eea
where $N$ denotes the nucleon, and in the second line we write in the mass basis. 
We take $\lambda_{hNN} = 2 \times 10^{-3}$~\cite{Shifman:1978zn,Bertone:2004pz}, 
but recent updates indicate a smaller value of $\lambda_{hNN} \approx 1.1 \times 10^{-3}$~\cite{Cheng:2012qr}.
We derive $\lambda_{\phi N N}$ using the formalism and notation of App.~C of Ref.~\cite{Bertone:2004pz}, 
to get the singlet VLQ (up-type $U$) contribution to the $\phi N N$ coupling via its contribution to the 
$\phi g g$ couplings, and the gluon content of the nucleon, which leads us to
$\lambda_{\phi NN} = (2/27)\, f^{(p,n)}_{TG}\, y_U m_{(p,n)}/M_U  \approx 0.063\, y_U\, m_N/M_U$. 
We can now write the $\psi$ elastic scattering cross section on a nucleon for $q^2 \ll m_N^2$ as 
\bea
\sigma(\psi N \to \psi N) &=& \frac{y_\psi^2}{8 \pi} 
   \left[\frac{s_h (c_h \lambda_{hNN} - s_h \lambda_{\phi NN})}{M_h^2} - \frac{c_h (c_h \lambda_{\phi NN} + s_h \lambda_{h NN})}{M_\phi^2}  \right]^2 
    \left( |{\bf p}_\psi|^2 + m_N^2 \right) \ , \nonumber \\
  &=& \frac{y_\psi^2 s_h^2 c_h^2 \lambda_{hNN}^2}{8 \pi} \frac{\left( |{\bf p}_\psi|^2 + m_N^2 \right)}{M_h^4} 
    \left[1 - \frac{\lambda_{\phi NN}}{\lambda_{h NN}} \frac{c_h}{s_h} \frac{(1+\Delta_\phi)}{(1-\Delta_h)} \frac{M_h^2}{M_\phi^2} \right]^2 \ ,
\label{dirDetCS.EQ}
\eea
where $p_\psi\approx M_\psi v_\psi$ with $v_\psi \sim 10^{-3}$~\cite{Bertone:2004pz}, $m_N \approx 1~$GeV is the nucleon mass, 
$\Delta_h = (\lambda_{\phi NN} / \lambda_{h NN}) (s_h/c_h)$, and $\Delta_\phi = (\lambda_{h NN} / \lambda_{\phi NN}) (s_h/c_h)$.
This is the generalization of the direct detection elastic cross section Eq.~(13) of Ref.~\cite{Gopalakrishna:2009yz}
which included only the $h$ contribution, to now include the $\phi$ contribution also that becomes important for very small $s_h$. 
\footnote{
For $s_h = 0.01$, the extra factor in Eq.~(\ref{dirDetCS.EQ}), namely, 
$\left[...\right]^2 \approx \left[1-0.3\, (1000~{\rm GeV}/M_U) (y_U/5) \right]^2$, with $\Delta_{\phi, h} \ll 1$ and can be dropped. 
Thus, for $s_h = 0.01$, $y_U = 5$, $M_U = 1000~$GeV, 
including the $\phi$ contribution {\em decreases} the elastic cross-section to about a half.
}
There is also uncertainty on the local dark matter halo density and its velocity distribution (for a discussion of these uncertainties, see for example Refs.~\cite{SigmaDD.Un}).
Given these uncertainties, our direct-detection rates should be taken to be accurate only up to unknown ${\cal O}(1)$ factors. 

\begin{figure}
\centering
\includegraphics[width=0.32\textwidth]{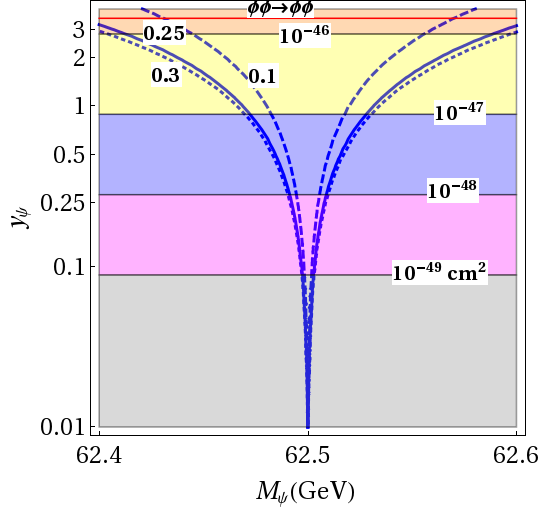}
\includegraphics[width=0.32\textwidth]{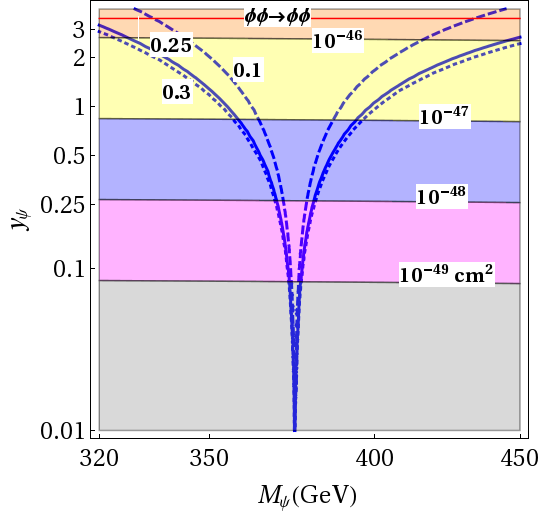}
\includegraphics[width=0.32\textwidth]{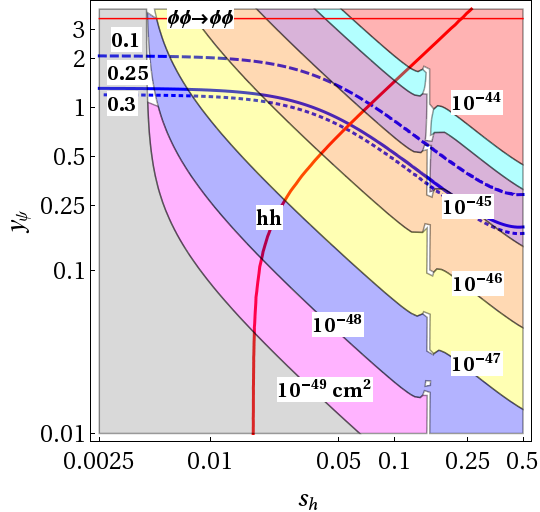}
\caption{In the $SVU\psi$ model for $Y_\psi = 0$, $Y_U = 2/3$,
  contours of $\Omega_{dm} = 0.1,~0.25,~0.3$,
  with the colored bands showing $\sigma_{DD}$ as marked, 
  for $y_U=5$, $M_U = 1000~$GeV, 
  and with the parameters not varied along the axes fixed at $s_h = 0.01$ and $M_\psi =350$~GeV.
  The red horizontal line shows the unitarity constraint
  for $y_U=5$, 
  and the thick red line shows the 8~TeV LHC $hh$ channel constraint.
}
\label{omega.svupsi}
\end{figure}

In Fig.~\ref{omega.svupsi} we plot contours of $\Omega_{dm} = 0.1,~0.25,~0.3$
and for $\lambda_N=2 \times 10^{-3}$, $m_N =1~$GeV,
show the regions with
$ \sigma_{DD} >  5*10^{-45}~$cm$^2$,
$10^{-45}~cm^2 < \sigma_{DD} < 5* 10^{-45}~$cm$^2$,
$ 10^{-46}~cm^2 < \sigma_{DD} < 10^{-45}~$cm$^2$,
$ 10^{-47}~cm^2 < \sigma_{DD} < 10^{-46}~$cm$^2$,
$ 10^{-48}~cm^2 < \sigma_{DD} < 10^{-47}~$cm$^2$,
$ 10^{-49}~cm^2 < \sigma_{DD} < 10^{-48}~$cm$^2$,
$ \sigma_{DD} < 10^{-49}~$cm$^2$
with parameters not varied along the axes fixed at $s_h=0.01$, 
$M_\psi = 350$~GeV, $y_U=5$ and $M_U = 1000$~GeV.
We also show in Fig.~\ref{omega.svupsi} the unitarity constrain on $y_\psi$
from $\phi \phi \to \phi \phi$ process for $y_U = 5$. 
We see that for $s_h =0.01$, $y_\psi \leq 4$ the direct detection cross section
is at or less than the current experimental limit $\sigma_{DD} \leq 10^{-45}$~cm$^2$~\cite{DirDetExptCite}.
For these values of $y_\psi$, the correct self-annihilation cross-section is obtained 
only with an enhancement of the cross-section at the $\phi,h$ pole with 
$M_\psi \sim M_{\phi,h}/2$.
Being close to the $\phi$ pole suppresses the $\phi \to \psi\psi$ decay rate due to the limited phase-space available, leading to a small $\kappa_\Gamma^2 \ll 0.1$ as can be seen from Fig.~\ref{sigmabr.svupsi}.

%%%%%%%%%%%%%%%%%%%%%%%%%%%%%%%%%%%%%%%%%%%%%%%%%%%%%%%%%%%%%%%%%%%%%%%%%%%
\section{Discussion and conclusions}
\label{DisCon.SEC}
In this work, we study the possibility that a scalar ($\phi$) with mass $750~$GeV explains the diphoton excess 
reported by the ATLAS and CMS experiments.
We ascertain the values the loop-induced $g g \phi$ and $\gamma\gamma \phi$ effective couplings must take in order to 
explain the observed diphoton cross section for various $\kappa_\Gamma^2 \equiv \Gamma_\phi/M_\phi$.
This is shown in Fig.~\ref{kggkaaAl}, which applies model-independently. 
A general observation is that for large $\Gamma_\phi$, obtaining the required diphoton rate needs large 
values of the effective couplings. 
Obtaining large effective couplings in a model will require some large coupling in it, which may violate perturbative unitarity.
We determine the upper limit on $\phi$-fermion-fermion couplings from requiring perturbative unitarity. 

We discuss two $SU(2)$ representation possibilities for $\phi$, namely the doublet and singlet.
We include the effects of standard model fermions (SMF), and vector-like fermions (VLF), in particular vector-like quarks (VLQ) and/or vector-like leptons (VLL) 
coupled to the $\phi$.   
These singlet or doublet scalars coupled to the SMF and/or VLF that we deal with here
can be thought of as extracts from various BSM constructions, that are relevant to explain the diphoton rate.
The VLF contributions are crucial to generate the required cross-section, especially if $\Gamma_\phi$ is large.

In the two-Higgs-doublet model (2HDM)
with the CP-odd and CP-even scalars $A,H$ taken to be at $M_A = 750~$GeV and $H$ lighter by 15~GeV, 
we explore in turn Type-I, Type-II and Type-X SMF couplings.
We find regions of parameter space consistent with $\phi\phi \to \phi\phi$ and $\psi\psi\to\psi\psi$ unitarity bounds,
and also with respect to direct 8~TeV LHC $t\bar t$ and $\tau\bar\tau$ limits.
A summary of the diphoton rate and total width achieved while satisfying the above constraints follows. 
We consider first a 2HDM with SMF only, and then with VLFs also present.
In Type-II 2HDM with only SMFs, the required diphoton rate cannot be obtained as we see from Fig.~\ref{sigmabr-2HDM.SMF}, 
and $\sigma_\phi \times BR_{\gamma\gamma}$ of only about $0.004~$fb can be reached
for a total width $16 \pi (\Gamma_\phi/M_\phi) \equiv \kappa_\Gamma^2 = 3$ (i.e. $\Gamma_\phi = 45~$GeV), which happens for $\tan\beta \approx 0.8$.
Adding VLL with a mass of $375~$GeV with SM-like hypercharge improves the situation significantly as we see from Fig.~\ref{sigmabr-2HDM.VLL},
but a diphoton rate of $0.5~$fb and $\kappa_\Gamma^2 \approx 0.25$ (i.e. $\Gamma_\phi = 3.7~$GeV) can be reached for Type-II couplings,
and $10~$fb and $\kappa_\Gamma^2 \approx 0.003$ (i.e. $\Gamma_\phi = 0.04~$GeV) can be reached for Type-I couplings.
By additionally adding VLQ with mass of $1000~$GeV, as seen from Fig.~\ref{sigmabr-2HDM.VLLQ}, a diphoton rate of about $10~$fb can be obtained
for a total width of about $\kappa_\Gamma^2 \approx 0.5$ (i.e. $\Gamma_\phi \approx 7.5~$GeV). 
A larger total width of $\kappa_\Gamma^2 \approx 2$ (i.e. $\Gamma_\phi \approx 30~$GeV) is possible, but only for a reduced diphoton rate of about $4~$fb. 

A singlet scalar $\phi$ cannot couple to SMF; we introduce SU(2) singlet VLL ($\psi$) and VLQ (EM charge +2/3 $U$) and couple it to $\phi$. 
We consider the two possibilities when the singlet VLL is charged and when it is neutral. 
The latter case gives the possibility of the neutral singlet VLL is a (hidden sector) dark matter, coupled to the SM sector
via Higgs singlet mixing generated after EWSB (i.e. the Higgs portal).
Another possibility of coupling the hidden sector VLL to SM is via the $\phi gg$ effective coupling induced by VLQ.
We explore this possibility also which is not usually included in the literature.
We introduce couplings between the $\phi$ and VLF and find regions of parameter-space that are 
compatible with respect to perturbative unitarity in the $\phi\phi\to\phi\phi$ and $\psi\psi\to\psi\psi$ channels, 
and 8~TeV LHC $hh$ channel constraint which restricts the size of the Higgs-singlet mixing. 
We also find regions of model parameter-space which give the correct dark matter relic density and dark matter direct detection.
All of these are shown in Fig.~\ref{omega.svupsi}.
As we see in this figure, obtaining the observed relic density requires $300$~GeV $\lesssim M_\psi \lesssim 450~$GeV (or $M_\psi \approx m_h/2$) to have a sufficiently large self-annihilation cross section
which is obtained only by hitting the $\phi$ (or $h$) pole in the s-channel.
We also find regions that explain the diphoton rate, with $M_\phi = 750~$GeV, 
with $\Gamma_\phi$ varied as shown in Figs.~\ref{sigmabr.svupsi}~and~\ref{sigmabr.svul}, 
the first for a neutral VLL, and the second for a VLL of EM charge $-1$.  
We find in these figures that a large diphoton rate as required is possible 
but only when $M_\psi \geq M_\phi/2$ for which $\kappa_\Gamma^2 < 0.1$ (i.e. $\Gamma_\phi < 1.5~$GeV). 
When $M_\psi < M_\phi/2$, the decay $\phi \to \psi\psi$ enhances the total width going even up to 
$\kappa_\Gamma^2 \lesssim 2$ (i.e. $\Gamma_\phi \lesssim 30~$GeV),
but causes a corresponding drop in the diphoton rate. 
For a total width so large, the diphoton rate can be large enough (i.e. a few fb value) only if $U$ is as light as $500~$GeV, 
which is not compatible with the LHC direct bounds unless it is so long-lived that it does not decay promptly and exits the detector.

In general, we observe that if the 750~GeV resonance is a scalar in the class of models we have considered,
the required diphoton rate can be obtained only by the addition of VLQ whose mass is not too much above a 1000~GeV. 
The direct search of the VLQ in this mass range at the LHC is perhaps the best way to test this hypothesis.
A study in this direction, although in a different context, is in Refs.~\cite{Gopalakrishna:2011ef,Gopalakrishna:2013hua} for example.
The addition of charged VLL with a mass of about $375~$GeV helps boost the $\phi\to \gamma\gamma$ partial width as we have seen, and is important to test. 
If a VLL $\psi$ with $M_\psi < M_\phi/2$ is present, $\phi \to\psi\psi$ decays are present and could be the reason for the large width of the $\phi$.
If $\psi$ decays promptly into some SM final state it may be observable at the LHC,
or if the decay is not prompt and $\phi$ has EM charge, may leave either a displaced vertex or a charged-track in the LHC detector, 
or, as we explored in detail in the singlet model, if $\psi$ is EM neutral and if it is stable over cosmological time-scales, it could be dark matter and can be searched for in dark matter direct detection experiments.
Yet another promising mode to look for at the LHC is the $\phi\to h h$ mode which already imposes very tight constraints on the parameter-space, although the size of this coupling is model dependent;
models in which $BR(\phi \to h h) \gtrsim 0.05$ may already be ruled out by the diHiggs LHC constraints. 
If $\phi$ is in the 2HDM, the charged Higgs search at the LHC becomes important.
Thus, the upcoming 13~TeV LHC run-II and dark matter direct-detection experiments may give us vital clues to test such models.

%%%%%%%%%%%%%%%%%%%%%%%%%
\medskip
\noindent {\it Acknowledgments:} 
We thank Rohini Godbole, Sreerup Raichaudhuri and Narendra Sahu for valuable discussions.

%%%%%%%%%%%%%%%%%%%%%%%%%%%%%%%%%%%%%%%%%%%%%%%%%%%%%%%%%%%%%%%%%%%%%%%%%%%
%\appendix
%%%%%%%%%%%%%%%%%%%%%%%%%%%%%%
%%%%%%%%%%%%%%%%%%%%%%%%%%%%%%%%%%%%%%%%%%%%%
% Put the following in place of the \appendix command
%%%%%%%%%%%%%%%%%%%%%%%%%%%%%%%%%%%%%%%%%%%%%
\setcounter{section}{0}
\renewcommand\thesection{\Alph{section}}               % use this to modify the section numbering style
\renewcommand\thesubsection{\Alph{section}.\arabic{subsection}}
\renewcommand\thesubsubsection{\Alph{section}.\arabic{subsection}.\arabic{subsubsection}}

\renewcommand{\theequation}{\Alph{section}.\arabic{equation}}    % use this to modify the equation numbers
\renewcommand{\thetable}{\Alph{section}.\arabic{table}}          % use this to modify the table numbers
\renewcommand{\thefigure}{\Alph{section}.\arabic{figure}}        % use this to modify the figure numbers

%%%%%%%%%%%%%%%%%%%%%%%%%%%%%%%%%%%%%%%%%%%%%%%%%%%
\section{Hidden sector dark matter relic density}
\label{hidRelDen.APP}

Here we give some details on the relic density calculation in the model of Sec.~\ref{singPhiHidDM.SEC}. 
The $|\mathcal{B}_i|^2$ for each of these final states are extracted from Ref.~\cite{Gopalakrishna:2009yz} to which we add $|\mathcal{B}_{gg}|^2$ here.
These are given by
\bea
|\mathcal{B}_{f\bar f}|^2&=&  N_c^f y_f^2 y_\psi^2 s_h^2 c_h^2 \left(1-\frac{4m_i^2}{s}\right) M_\psi^4 \hat{S}_{BW}^{h\phi} \ ; \qquad
\hat{S}_{BW}^{h\phi} = \frac{(M_\phi^2 - M_h^2)^2}{\left[ (s-M_h^2)^2 + M_h^2 \Gamma_h^2 \right] \left[ (s-M_\phi^2)^2 + M_\phi^2 \Gamma_\phi^2  \right]  } \ , \nonumber \\
|\mathcal{B}_{WW}|^2 &=& \frac{1}{4} y_\psi^2 g^4 v^2 s_h^2 c_h^2 M_\psi^2 \left[ \frac{1}{2} + \frac{(s/2 - M_W^2)^2}{4 M_W^4} \right] \hat{S}_{BW}^{h\phi} \ , \\
|\mathcal{B}_{hh}|^2 &=& \frac{ M_\psi^2 y_\psi^2}{64} \left\{ \frac{ s_h^2 c_h^6 \kappa_{3h}^2 v^2}{\left[ (s-M_h^2)^2 + M_h^2 \Gamma_h^2 \right]} + \frac{c_h^8 \kappa_{\phi hh}^2 M_\phi^2}{\left[ (s-M_\phi^2)^2 + M_\phi^2 \Gamma_\phi^2 \right]}
  - \frac{2  s_h c_h^7 \kappa_{3h} v \kappa_{\phi hh} M_\phi}{\left[ (s-M_h^2) (s-M_\phi^2) + M_h M_\phi \Gamma_h \Gamma_\phi \right] } \right\} \nonumber \ , \nonumber \\
  |\mathcal{B}_{gg}|^2 &=&  \frac{16 y_\psi^2 M_\psi^6}{(16 \pi^2 M)^2}  \left\{\frac{c_h^2 \kappa_{\phi gg}^2}{(s- M_\phi^2)^2 + M_\phi^2 \Gamma_\phi^2}    + \frac{s_h^2 \kappa_{h gg}^2}{(s- M_h^2)^2 + M_h^2 \Gamma_h^2} -\frac{2 c_h s_h \kappa_{\phi gg} \kappa_{h gg} }{\left[ (s-M_h^2) (s-M_\phi^2) + M_h M_\phi \Gamma_h \Gamma_\phi \right]}\right\}  \nonumber
\eea
where $s\approx 4 M_\psi^2$, $\hat{S}_{BW}^{h\phi}$ is a Breit-Wigner resonance factor including the s-channel $\{h,\phi\}$ contributions, $f\bar f = \{ b\bar b, t\bar t\}$,
the ${\cal M}_{ZZ}$ is identical to ${\cal M}_{WW}$ except for an additional factor of $1/(2c_W^2)$ and $M_W \to M_Z$,
and in $|{\cal M}_{hh}|$ we do not include the t-channel contributions as these are sub-dominant;
$M$ is a mass scale which we set to $1$~TeV for numerical evaluations and
the mixing angle $\theta_h$ enters in $\kappa_{\phi gg}$ and $\kappa_{h gg}$ through $\phi UU, \phi tt$ and $htt$ couplings.

%%%%%%%%%%%%%%%%%%%%%%%%%%%%%%%%%%%%%%%%%%


\begin{thebibliography}{99}

%\cite{ATLAS-750GeVExcess}
\bibitem{ATLAS-750GeVExcess} 
  The ATLAS collaboration,
  %``Search for resonances decaying to photon pairs in 3.2 fb$^{-1}$ of $pp$ collisions at $\sqrt{s}$ = 13 TeV with the ATLAS detector,''
  ATLAS-CONF-2015-081.
  %%CITATION = ATLAS-CONF-2015-081;%%
  %267 citations counted in INSPIRE as of 25 Mar 2016

\bibitem{CMS-750GeVExcess}
%\cite{CMS:2015dxe}
%\bibitem{CMS:2015dxe} 
  CMS Collaboration [CMS Collaboration],
  %``Search for new physics in high mass diphoton events in proton-proton collisions at 13TeV,''
  CMS-PAS-EXO-15-004;
  %%CITATION = CMS-PAS-EXO-15-004;%%
  %259 citations counted in INSPIRE as of 25 Mar 2016
  %
  
  %\cite{Franceschini:2015kwy}
\bibitem{Franceschini:2015kwy} 
  R.~Franceschini {\it et al.},
  %``What is the $\gamma \gamma$ resonance at 750 GeV?,''
  JHEP {\bf 1603}, 144 (2016)
  doi:10.1007/JHEP03(2016)144
  [arXiv:1512.04933 [hep-ph]].
  %%CITATION = doi:10.1007/JHEP03(2016)144;%%
  %236 citations counted in INSPIRE as of 04 Apr 2016owr
  %
%\cite{Gupta:2015zzs}
\bibitem{Gupta:2015zzs} 
  R.~S.~Gupta, S.~Jäger, Y.~Kats, G.~Perez and E.~Stamou,
  %``Interpreting a 750 GeV Diphoton Resonance,''
  arXiv:1512.05332 [hep-ph].
  %%CITATION = ARXIV:1512.05332;%%
  %176 citations counted in INSPIRE as of 04 Apr 2016owr
  %
  
%\cite{CMS:2016owr}
%\bibitem{CMS:2016owr} 
  CMS Collaboration [CMS Collaboration],
  %``Search for new physics in high mass diphoton events in $3.3~\mathrm{fb}^{-1}$ of proton-proton collisions at $\sqrt{s}=13~\mathrm{TeV}$ and combined interpretation of searches at $8~\mathrm{TeV}$ and $13~\mathrm{TeV}$,''
  CMS-PAS-EXO-16-018.
  %%CITATION = CMS-PAS-EXO-16-018;%%
  %2 citations counted in INSPIRE as of 25 Mar 2016


  %%%%%%%%%%%%%%%%%%%%%%%%%%%%%%%%%%%%%%%%%%%%%%%%%%%%%%%%%%%%%%%%%%%%%%%%%%%%%%%%%%%%%%%%%%%%%%%%%%%%%%%%%%%%%%%%%%%%%
  %\cite{Djouadi:2016eyy}
\bibitem{Djouadi:2016eyy} 
  A.~Djouadi, J.~Ellis, R.~Godbole and J.~Quevillon,
  %``Future Collider Signatures of the Possible 750 GeV State,''
  arXiv:1601.03696 [hep-ph].
  %%CITATION = ARXIV:1601.03696;%%
  %40 citations counted in INSPIRE as of 05 Apr 2016dez
  %
  
  %\cite{Badziak:2015zez}
\bibitem{Badziak:2015zez} 
  M.~Badziak,
  %``Interpreting the 750 GeV diphoton excess in minimal extensions of Two-Higgs-Doublet models,''
  arXiv:1512.07497 [hep-ph].
  %%CITATION = ARXIV:1512.07497;%%
  %74 citations counted in INSPIRE as of 04 Apr 2016
%

%\cite{Bizot:2015qqo}
\bibitem{Bizot:2015qqo} 
  N.~Bizot, S.~Davidson, M.~Frigerio and J.-L.~Kneur,
  %``Two Higgs doublets to explain the excesses $pp\rightarrow \gamma\gamma(750\ {\rm GeV})$ and $h \to \tau^\pm \mu^\mp$,''
  JHEP {\bf 1603}, 073 (2016)
  doi:10.1007/JHEP03(2016)073
  [arXiv:1512.08508 [hep-ph]].
  %%CITATION = doi:10.1007/JHEP03(2016)073;%%
  %69 citations counted in INSPIRE as of 05 Apr 2016owr
  %
  
  %\cite{Bertuzzo:2016fmv}
\bibitem{Bertuzzo:2016fmv} 
  E.~Bertuzzo, P.~A.~N.~Machado and M.~Taoso,
  %``Di-Photon excess in the 2HDM: hasting towards the instability and the non-perturbative regime,''
  arXiv:1601.07508 [hep-ph].
  %%CITATION = ARXIV:1601.07508;%%
  %15 citations counted in INSPIRE as of 05 Apr 2016fmv
  %
  
  %\cite{Bharucha:2016jyr}
\bibitem{Bharucha:2016jyr} 
  A.~Bharucha, A.~Djouadi and A.~Goudelis,
  %``Threshold enhancement of diphoton resonances,''
  arXiv:1603.04464 [hep-ph].
  %%CITATION = ARXIV:1603.04464;%%
  %3 citations counted in INSPIRE as of 05 Apr 2016fmv
  %
  
  %\cite{Angelescu:2015uiz}
\bibitem{Angelescu:2015uiz} 
  A.~Angelescu, A.~Djouadi and G.~Moreau,
  %``Scenarii for interpretations of the LHC diphoton excess: two Higgs doublets and vector-like quarks and leptons,''
  Phys.\ Lett.\ B {\bf 756}, 126 (2016)
  doi:10.1016/j.physletb.2016.02.064
  [arXiv:1512.04921 [hep-ph]].
  %%CITATION = doi:10.1016/j.physletb.2016.02.064;%%
  %166 citations counted in INSPIRE as of 05 Apr 2016
  
  
  %%%%%%%%%%%%%%%%%%%%%%%%%%%%%%%%%%%%%%%%%%%%%%%%%%%%%%%%%%%%%%%%%%%%%%%%%%%%%%%%%%%%%%%%%%%%%%%%%%%%%%%%%%%%%%%%5
    
  
  %\cite{DiChiara:2015vdm}
\bibitem{DiChiara:2015vdm} 
  S.~Di Chiara, L.~Marzola and M.~Raidal,
  %``First interpretation of the 750 GeV di-photon resonance at the LHC,''
  arXiv:1512.04939 [hep-ph].
  %%CITATION = ARXIV:1512.04939;%%
  %162 citations counted in INSPIRE as of 04 Apr 2016owr
  %
  
  %\cite{McDermott:2015sck}
\bibitem{McDermott:2015sck} 
  S.~D.~McDermott, P.~Meade and H.~Ramani,
  %``Singlet Scalar Resonances and the Diphoton Excess,''
  Phys.\ Lett.\ B {\bf 755}, 353 (2016)
  doi:10.1016/j.physletb.2016.02.033
  [arXiv:1512.05326 [hep-ph]].
  %%CITATION = doi:10.1016/j.physletb.2016.02.033;%%
  %164 citations counted in INSPIRE as of 04 Apr 2016
%

%\cite{Benbrik:2015fyz}
\bibitem{Benbrik:2015fyz} 
  R.~Benbrik, C.~H.~Chen and T.~Nomura,
  %``Higgs singlet boson as a diphoton resonance in a vectorlike quark model,''
  Phys.\ Rev.\ D {\bf 93}, no. 5, 055034 (2016)
  doi:10.1103/PhysRevD.93.055034
  [arXiv:1512.06028 [hep-ph]].
  %%CITATION = doi:10.1103/PhysRevD.93.055034;%%
  %104 citations counted in INSPIRE as of 04 Apr 2016
  
%\cite{Ellis:2015oso}
\bibitem{Ellis:2015oso} 
  J.~Ellis, S.~A.~R.~Ellis, J.~Quevillon, V.~Sanz and T.~You,
  %``On the Interpretation of a Possible $\sim 750$ GeV Particle Decaying into $\gamma \gamma$,''
  JHEP {\bf 1603}, 176 (2016)
  doi:10.1007/JHEP03(2016)176
  [arXiv:1512.05327 [hep-ph]].
  %%CITATION = doi:10.1007/JHEP03(2016)176;%%
  %184 citations counted in INSPIRE as of 04 Apr 2016owr
  %
  
  %\cite{Curtin:2015jcv}
\bibitem{Curtin:2015jcv} 
  D.~Curtin and C.~B.~Verhaaren,
  %``Quirky Explanations for the Diphoton Excess,''
  Phys.\ Rev.\ D {\bf 93}, no. 5, 055011 (2016)
  doi:10.1103/PhysRevD.93.055011
  [arXiv:1512.05753 [hep-ph]].
  %%CITATION = doi:10.1103/PhysRevD.93.055011;%%
  %131 citations counted in INSPIRE as of 04 Apr 2016owr
  %
  
  %\cite{Falkowski:2015swt}
\bibitem{Falkowski:2015swt} 
  A.~Falkowski, O.~Slone and T.~Volansky,
  %``Phenomenology of a 750 GeV Singlet,''
  JHEP {\bf 1602}, 152 (2016)
  doi:10.1007/JHEP02(2016)152
  [arXiv:1512.05777 [hep-ph]].
  %%CITATION = doi:10.1007/JHEP02(2016)152;%%
  %177 citations counted in INSPIRE as of 04 Apr 2016owr
  %
  
  %\cite{Aloni:2015mxa}
\bibitem{Aloni:2015mxa} 
  D.~Aloni, K.~Blum, A.~Dery, A.~Efrati and Y.~Nir,
  %``On a possible large width 750 GeV diphoton resonance at ATLAS and CMS,''
  arXiv:1512.05778 [hep-ph].
  %%CITATION = ARXIV:1512.05778;%%
  %120 citations counted in INSPIRE as of 04 Apr 2016owr
  %
  
  %\cite{Cheung:2015cug}
\bibitem{Cheung:2015cug} 
  K.~Cheung, P.~Ko, J.~S.~Lee, J.~Park and P.~Y.~Tseng,
  %``A Higgcision study on the 750 GeV Di-photon Resonance and 125 GeV SM Higgs boson with the Higgs-Singlet Mixing,''
  arXiv:1512.07853 [hep-ph].
  %%CITATION = ARXIV:1512.07853;%%
  %78 citations counted in INSPIRE as of 04 Apr 2016owr
  %
  
  %\cite{Bhattacharya:2016lyg}
\bibitem{Bhattacharya:2016lyg} 
  S.~Bhattacharya, S.~Patra, N.~Sahoo and N.~Sahu,
  %``750 GeV Di-Photon Excess at CERN LHC from a Dark Sector Assisted Scalar Decay,''
  arXiv:1601.01569 [hep-ph].
  %%CITATION = ARXIV:1601.01569;%%
  %38 citations counted in INSPIRE as of 05 Apr 2016lyg
  %
  
  %\cite{D'Eramo:2016mgv}
\bibitem{D'Eramo:2016mgv} 
  F.~D'Eramo, J.~de Vries and P.~Panci,
  %``A 750 GeV Portal: LHC Phenomenology and Dark Matter Candidates,''
  arXiv:1601.01571 [hep-ph].
  %%CITATION = ARXIV:1601.01571;%%
  %41 citations counted in INSPIRE as of 05 Apr 2016lyg
  %
  
  %\cite{Kawamura:2016idj}
\bibitem{Kawamura:2016idj} 
  J.~Kawamura and Y.~Omura,
  %``Diphoton excess at 750 GeV and LHC constraints in models with vector-like particles,''
  arXiv:1601.07396 [hep-ph].
  %%CITATION = ARXIV:1601.07396;%%
  %17 citations counted in INSPIRE as of 05 Apr 2016idj
  %
  
  %\cite{Chen:2016sck}
\bibitem{Chen:2016sck} 
  C.~Y.~Chen, M.~Lefebvre, M.~Pospelov and Y.~M.~Zhong,
  %``Diphoton Excess through Dark Mediators,''
  arXiv:1603.01256 [hep-ph].
  %%CITATION = ARXIV:1603.01256;%%
  %3 citations counted in INSPIRE as of 05 Apr 2016fmv
  %
  
  %\cite{DiChiara:2016dez}
\bibitem{DiChiara:2016dez} 
  S.~Di Chiara, A.~Hektor, K.~Kannike, L.~Marzola and M.~Raidal,
  %``Large loop-coupling enhancement of a 750 GeV pseudoscalar from a light dark sector,''
  arXiv:1603.07263 [hep-ph].
  %%CITATION = ARXIV:1603.07263;%%
  %1 citations counted in INSPIRE as of 05 Apr 2016dez
  %
  
   %\cite{Mambrini:2015wyu}
\bibitem{Mambrini:2015wyu} 
  Y.~Mambrini, G.~Arcadi and A.~Djouadi,
  %``The LHC diphoton resonance and dark matter,''
  Phys.\ Lett.\ B {\bf 755}, 426 (2016)
  doi:10.1016/j.physletb.2016.02.049
  [arXiv:1512.04913 [hep-ph]].
  %%CITATION = doi:10.1016/j.physletb.2016.02.049;%%
  %169 citations counted in INSPIRE as of 05 avril 2016dez
  %
  
  %\cite{Kanemura:2015vcb}
\bibitem{Kanemura:2015vcb} 
  S.~Kanemura, N.~Machida, S.~Odori and T.~Shindou,
  %``Diphoton excess at 750 GeV in an extended scalar sector,''
  arXiv:1512.09053 [hep-ph].
  %%CITATION = ARXIV:1512.09053;%%
  %48 citations counted in INSPIRE as of 12 Apr 2016dez
  %
  
  %\cite{DeRomeri:2016xpb}
\bibitem{DeRomeri:2016xpb} 
  V.~De Romeri, J.~S.~Kim, V.~Martin-Lozano, K.~Rolbiecki and R.~R.~de Austri,
  %``Confronting dark matter with the diphoton excess from a parent resonance decay,''
  arXiv:1603.04479 [hep-ph].
  %%CITATION = ARXIV:1603.04479;%%
  %4 citations counted in INSPIRE as of 12 Apr 2016dez
  %
  
  %\cite{Hamada:2016vwk}
\bibitem{Hamada:2016vwk} 
  Y.~Hamada, H.~Kawai, K.~Kawana and K.~Tsumura,
  %``Models of LHC Diphoton Excesses Valid up to the Planck scale,''
  arXiv:1602.04170 [hep-ph].
  %%CITATION = ARXIV:1602.04170;%%
  %16 citations counted in INSPIRE as of 12 Apr 2016dez
  %
  
  %\cite{Godunov:2016kqn}
\bibitem{Godunov:2016kqn} 
  S.~I.~Godunov, A.~N.~Rozanov, M.~I.~Vysotsky and E.~V.~Zhemchugov,
  %``New Physics at 1 TeV?,''
  arXiv:1602.02380 [hep-ph].
  %%CITATION = ARXIV:1602.02380;%%
  %9 citations counted in INSPIRE as of 12 Apr 2010ae
  %
  
  %\cite{Ge:2016xcq}
\bibitem{Ge:2016xcq} 
  S.~F.~Ge, H.~J.~He, J.~Ren and Z.~Z.~Xianyu,
  %``Realizing Dark Matter and Higgs Inflation in Light of LHC Diphoton Excess,''
  arXiv:1602.01801 [hep-ph].
  %%CITATION = ARXIV:1602.01801;%%
  %20 citations counted in INSPIRE as of 12 Apr 2016dez%
  %
  %\cite{Ko:2016wce}
\bibitem{Ko:2016wce} 
  P.~Ko and T.~Nomura,
  %``Dark sector shining through 750 GeV dark Higgs boson at the LHC,''
  arXiv:1601.02490 [hep-ph].
  %%CITATION = ARXIV:1601.02490;%%
  %33 citations counted in INSPIRE as of 12 Apr 2016dez
  %
  
  %\cite{Cai:2015hzc}
\bibitem{Cai:2015hzc} 
  C.~Cai, Z.~H.~Yu and H.~H.~Zhang,
  %``The 750 GeV diphoton resonance as a singlet scalar in an extra dimensional model,''
  arXiv:1512.08440 [hep-ph].
  %%CITATION = ARXIV:1512.08440;%%
  %65 citations counted in INSPIRE as of 12 Apr 2016dez
  %
  
%\cite{Zhang:2015uuo}
\bibitem{Zhang:2015uuo} 
  J.~Zhang and S.~Zhou,
  %``Electroweak Vacuum Stability and Diphoton Excess at 750 GeV,''
  arXiv:1512.07889 [hep-ph].
  %%CITATION = ARXIV:1512.07889;%%
  %81 citations counted in INSPIRE as of 12 Apr 2016dez
  %
  
  %\cite{Han:2015dlp}
\bibitem{Han:2015dlp} 
  H.~Han, S.~Wang and S.~Zheng,
  %``Scalar Explanation of Diphoton Excess at LHC,''
  arXiv:1512.06562 [hep-ph].
  %%CITATION = ARXIV:1512.06562;%%
  %99 citations counted in INSPIRE as of 12 Apr 2016dez
  %
  
  %\cite{Redi:2016kip}
\bibitem{Redi:2016kip} 
  M.~Redi, A.~Strumia, A.~Tesi and E.~Vigiani,
  %``Di-photon resonance and Dark Matter as heavy pions,''
  arXiv:1602.07297 [hep-ph].
  %%CITATION = ARXIV:1602.07297;%%
  %12 citations counted in INSPIRE as of 12 Apr 2016dez
  %
  
  %\cite{Chakraborty:2015jvs}
\bibitem{Chakraborty:2015jvs} 
  I.~Chakraborty and A.~Kundu,
  %``Diphoton excess at 750 GeV: Singlet scalars confront triviality,''
  Phys.\ Rev.\ D {\bf 93}, no. 5, 055003 (2016)
  doi:10.1103/PhysRevD.93.055003
  [arXiv:1512.06508 [hep-ph]].
  %%CITATION = doi:10.1103/PhysRevD.93.055003;%%
  %87 citations counted in INSPIRE as of 12 Apr 2016dez
  %
  
  %\cite{Backovic:2015fnp}
\bibitem{Backovic:2015fnp} 
  M.~Backovic, A.~Mariotti and D.~Redigolo,
  %``Di-photon excess illuminates Dark Matter,''
  JHEP {\bf 1603}, 157 (2016)
  doi:10.1007/JHEP03(2016)157
  [arXiv:1512.04917 [hep-ph]].
  %%CITATION = doi:10.1007/JHEP03(2016)157;%%
  %169 citations counted in INSPIRE as of 12 Apr 2016
  
  %%%%%%%%%%%%%%%%%%%%%%%%%%%%%%%%%%%%%%%%%%%%%%%%%%%%%%%%%%%%%%%%%%%%%%%%%%%%%%%%%%%%%%%%%%%%%%%%%%%%%%%%%%%%%%%%%%%%%%%%%%%%%%%%%%%%%%%%
  \bibitem{Baglio:2010ae} 
  J.~Baglio and A.~Djouadi,
  %``Higgs production at the lHC,''
  JHEP {\bf 1103}, 055 (2011)
  [arXiv:1012.0530 [hep-ph]].
  %%CITATION = ARXIV:1012.0530;%%
  %158 citations counted in INSPIRE as of 12 Sep 2014aqa
  %
  
  %\cite{Gopalakrishna:2015wwa}
\bibitem{Gopalakrishna:2015wwa} 
  S.~Gopalakrishna, T.~S.~Mukherjee and S.~Sadhukhan,
  %``Extra neutral scalars with vectorlike fermions at the LHC,''
  Phys.\ Rev.\ D {\bf 93}, no. 5, 055004 (2016)
  doi:10.1103/PhysRevD.93.055004
  [arXiv:1504.01074 [hep-ph]].
  %%CITATION = doi:10.1103/PhysRevD.93.055004;%%
  %4 citations counted in INSPIRE as of 25 Mar 2016  
  %
  
  %\cite{Gunion:1989we}
\bibitem{Gunion:1989we} 
  J.~F.~Gunion, H.~E.~Haber, G.~L.~Kane and S.~Dawson,
  %``The Higgs Hunter's Guide,''
  Front.\ Phys.\  {\bf 80}, 1 (2000).
  %%CITATION = FRPHA,80,1;%%
  %497 citations counted in INSPIRE as of 12 Jan 2016
  %
  
  %\cite{Aad:2015xja}
\bibitem{Aad:2015xja} 
  G.~Aad {\it et al.} [ATLAS Collaboration],
  %``Searches for Higgs boson pair production in the $hh\to bb\tau\tau, \gamma\gamma WW^*, \gamma\gamma bb, bbbb$ channels with the ATLAS detector,''
  Phys.\ Rev.\ D {\bf 92}, 092004 (2015)
  doi:10.1103/PhysRevD.92.092004
  [arXiv:1509.04670 [hep-ex]].
  %%CITATION = doi:10.1103/PhysRevD.92.092004;%%
  %34 citations counted in INSPIRE as of 25 Mar 2016

%\cite{CMS:2015neg}
\bibitem{CMS:2015neg} 
  CMS Collaboration [CMS Collaboration],
  %``Search for Resonances Decaying to Dijet Final States at $\sqrt{s} = 8$ TeV with Scouting Data,''
  CMS-PAS-EXO-14-005.
  %%CITATION = CMS-PAS-EXO-14-005;%%
  %47 citations counted in INSPIRE as of 25 Mar 2016 

%\cite{Aad:2014aqa}
\bibitem{Aad:2014aqa} 
  G.~Aad {\it et al.} [ATLAS Collaboration],
  %``Search for new phenomena in the dijet mass distribution using $p-p$ collision data at $\sqrt{s}=8$ TeV with the ATLAS detector,''
  Phys.\ Rev.\ D {\bf 91}, no. 5, 052007 (2015)
  doi:10.1103/PhysRevD.91.052007
  [arXiv:1407.1376 [hep-ex]].
  %%CITATION = doi:10.1103/PhysRevD.91.052007;%%
  %191 citations counted in INSPIRE as of 25 Mar 2016  
  
  %\cite{Ellis:2014dza}
\bibitem{Ellis:2014dza} 
  S.~A.~R.~Ellis, R.~M.~Godbole, S.~Gopalakrishna and J.~D.~Wells,
  %``Survey of vector-like fermion extensions of the Standard Model and their phenomenological implications,''
  JHEP {\bf 1409}, 130 (2014)
  doi:10.1007/JHEP09(2014)130
  [arXiv:1404.4398 [hep-ph]].
  %%CITATION = doi:10.1007/JHEP09(2014)130;%%
  %37 citations counted in INSPIRE as of 17 Feb 2016dez
  %

%\cite{ATLAS:2013ima}
\bibitem{ATLAS:2013ima} 
  [ATLAS Collaboration],
  %``Search for heavy top-like quarks decaying to a Higgs boson and a top quark in the lepton plus jets final state in $pp$ collisions at $\sqrt{s}=8$ TeV with the ATLAS detector,''
  ATLAS-CONF-2013-018.
  %%CITATION = ATLAS-CONF-2013-018;%%
  %80 citations counted in INSPIRE as of 23 Mar 2016  

%\cite{CMS:2013tda}
\bibitem{CMS:2013tda} 
  CMS Collaboration [CMS Collaboration],
  %``Inclusive search for a vector-like T quark by CMS,''
  CMS-PAS-B2G-12-015.
  %%CITATION = CMS-PAS-B2G-12-015;%%
  %27 citations counted in INSPIRE as of 23 Mar 2016;
  %\cite{Khachatryan:2015oba}
  
\bibitem{Khachatryan:2015oba} 
  V.~Khachatryan {\it et al.} [CMS Collaboration],
  %``Search for vector-like charge 2/3 T quarks in proton-proton collisions at sqrt(s) = 8 TeV,''
  Phys.\ Rev.\ D {\bf 93}, no. 1, 012003 (2016)
  doi:10.1103/PhysRevD.93.012003
  [arXiv:1509.04177 [hep-ex]].
  %%CITATION = doi:10.1103/PhysRevD.93.012003;%%
  %27 citations counted in INSPIRE as of 25 Mar 2016

%\cite{}
\bibitem{ATLAS:tp-13TeV-3.2ifb} 
  The ATLAS collaboration,
  %``Search for production of vector-like top quark pairs and of four top quarks in the lepton-plus-jets final state in $pp$ collisions at $\sqrt{s}=13$ TeV with the ATLAS detector,''
  ATLAS-CONF-2016-013.
  %%CITATION = ATLAS-CONF-2016-013;%%

%\cite{Aad:2015kqa}
\bibitem{Aad:2015kqa} 
  G.~Aad {\it et al.} [ATLAS Collaboration],
  %``Search for production of vector-like quark pairs and of four top quarks in the lepton-plus-jets final state in $pp$ collisions at $\sqrt{s}=8$ TeV with the ATLAS detector,''
  JHEP {\bf 1508}, 105 (2015)
  doi:10.1007/JHEP08(2015)105
  [arXiv:1505.04306 [hep-ex]].
  %%CITATION = doi:10.1007/JHEP08(2015)105;%%
  %75 citations counted in INSPIRE as of 25 Mar 2016
 
%\cite{Aad:2015mba}
\bibitem{Aad:2015mba} 
  G.~Aad {\it et al.} [ATLAS Collaboration],
  %``Search for vector-like $B$ quarks in events with one isolated lepton, missing transverse momentum and jets at $\sqrt{s}=$ 8 TeV with the ATLAS detector,''
  Phys.\ Rev.\ D {\bf 91}, no. 11, 112011 (2015)
  doi:10.1103/PhysRevD.91.112011
  [arXiv:1503.05425 [hep-ex]].
  %%CITATION = doi:10.1103/PhysRevD.91.112011;%%
  %35 citations counted in INSPIRE as of 25 Mar 2016

  %\cite{Khachatryan:2015gza}
\bibitem{Khachatryan:2015gza} 
  V.~Khachatryan {\it et al.} [CMS Collaboration],
  %``Search for pair-produced vector-like B quarks in proton-proton collisions at $\sqrt{s}$ = 8 TeV,''
  arXiv:1507.07129 [hep-ex].
  %%CITATION = ARXIV:1507.07129;%%
  %18 citations counted in INSPIRE as of 25 Mar 2016

%\cite{Khachatryan:2015jha}
\bibitem{Khachatryan:2015jha} 
  V.~Khachatryan {\it et al.} [CMS Collaboration],
  %``Search for Decays of Stopped Long-Lived Particles Produced in Proton–Proton Collisions at $\sqrt{s}= 8\,\text {TeV} $,''
  Eur.\ Phys.\ J.\ C {\bf 75}, no. 4, 151 (2015)
  doi:10.1140/epjc/s10052-015-3367-z
  [arXiv:1501.05603 [hep-ex]].
  %%CITATION = doi:10.1140/epjc/s10052-015-3367-z;%%
  %20 citations counted in INSPIRE as of 25 Mar 2016

%\cite{Aad:2013gva}
\bibitem{Aad:2013gva} 
  G.~Aad {\it et al.} [ATLAS Collaboration],
  %``Search for long-lived stopped R-hadrons decaying out-of-time with pp collisions using the ATLAS detector,''
  Phys.\ Rev.\ D {\bf 88}, no. 11, 112003 (2013)
  doi:10.1103/PhysRevD.88.112003
  [arXiv:1310.6584 [hep-ex]].
  %%CITATION = doi:10.1103/PhysRevD.88.112003;%%
  %52 citations counted in INSPIRE as of 25 Mar 2016
  
%\cite{Falkowski:2013jya}
\bibitem{Falkowski:2013jya} 
  A.~Falkowski, D.~M.~Straub and A.~Vicente,
  %``Vector-like leptons: Higgs decays and collider phenomenology,''
  JHEP {\bf 1405}, 092 (2014)
  doi:10.1007/JHEP05(2014)092
  [arXiv:1312.5329 [hep-ph]].
  %%CITATION = doi:10.1007/JHEP05(2014)092;%%
  %58 citations counted in INSPIRE as of 25 Mar 2016
  %
  
  %\cite{Agashe:2014kda}
\bibitem{Agashe:2014kda} 
  K.~A.~Olive {\it et al.} [Particle Data Group Collaboration],
  %``Review of Particle Physics,''
  Chin.\ Phys.\ C {\bf 38}, 090001 (2014),
  doi:10.1088/1674-1137/38/9/090001.
  %%CITATION = doi:10.1088/1674-1137/38/9/090001;%%
  %2355 citations counted in INSPIRE as of 23 Nov 2015

%\cite{Peskin:1995ev}
\bibitem{Peskin:1995ev} 
  M.~E.~Peskin and D.~V.~Schroeder,
  %``An Introduction to quantum field theory,''
  Reading, USA: Addison-Wesley (1995) 842 p
  %936 citations counted in INSPIRE as of 03 Mar 2016dez
  %
  
  %\cite{Branco:2011iw}
\bibitem{Branco:2011iw} 
  G.~C.~Branco, P.~M.~Ferreira, L.~Lavoura, M.~N.~Rebelo, M.~Sher and J.~P.~Silva,
  %``Theory and phenomenology of two-Higgs-doublet models,''
  Phys.\ Rept.\  {\bf 516}, 1 (2012)
  doi:10.1016/j.physrep.2012.02.002
  [arXiv:1106.0034 [hep-ph]].
  %%CITATION = doi:10.1016/j.physrep.2012.02.002;%%
  %666 citations counted in INSPIRE as of 18 Mar 2016dez
  %
  
  %\cite{Bernon:2015qea}
\bibitem{Bernon:2015qea} 
  J.~Bernon, J.~F.~Gunion, H.~E.~Haber, Y.~Jiang and S.~Kraml,
  %``Scrutinizing the alignment limit in two-Higgs-doublet models: m$_h$=125  GeV,''
  Phys.\ Rev.\ D {\bf 92}, no. 7, 075004 (2015)
  doi:10.1103/PhysRevD.92.075004
  [arXiv:1507.00933 [hep-ph]].
  %%CITATION = doi:10.1103/PhysRevD.92.075004;%%
  %15 citations counted in INSPIRE as of 18 Mar 2016
  
  %\cite{Bhattacharyya:2015nca}
\bibitem{Bhattacharyya:2015nca} 
  G.~Bhattacharyya and D.~Das,
  %``Scalar sector of Two-Higgs-Doublet models: A mini-review,''
  arXiv:1507.06424 [hep-ph].
  %%CITATION = ARXIV:1507.06424;%%
  %4 citations counted in INSPIRE as of 18 Mar 2016dez
  %
  
  %\cite{Gopalakrishna:2015dkt}
\bibitem{Gopalakrishna:2015dkt} 
  S.~Gopalakrishna, T.~S.~Mukherjee and S.~Sadhukhan,
  %``Status and Prospects of the Two-Higgs-Doublet SU(6)/Sp(6) little-Higgs Model and the Alignment Limit,''
  arXiv:1512.05731 [hep-ph].
  %%CITATION = ARXIV:1512.05731;%%
  %4 citations counted in INSPIRE as of 25 Mar 2016dez
  %
  
   %\cite{Djouadi:2005gj}
\bibitem{Djouadi:2005gj} 
  A.~Djouadi,
  %``The Anatomy of electro-weak symmetry breaking. II. The Higgs bosons in the minimal supersymmetric model,''
  Phys.\ Rept.\  {\bf 459}, 1 (2008)
  doi:10.1016/j.physrep.2007.10.005
  [hep-ph/0503173].
  %%CITATION = doi:10.1016/j.physrep.2007.10.005;%%
  %929 citations counted in INSPIRE as of 14 Mar 2016
  
  %\cite{Chang:2012ve}
\bibitem{Chang:2012ve} 
  S.~Chang, S.~K.~Kang, J.~P.~Lee, K.~Y.~Lee, S.~C.~Park and J.~Song,
  %``Comprehensive study of two Higgs doublet model in light of the new boson with mass around 125 GeV,''
  JHEP {\bf 1305}, 075 (2013)
  doi:10.1007/JHEP05(2013)075
  [arXiv:1210.3439 [hep-ph]].
  %%CITATION = doi:10.1007/JHEP05(2013)075;%%
  %67 citations counted in INSPIRE as of 14 Mar 2016dez
  %


%\cite{Gopalakrishna:2009yz}
\bibitem{Gopalakrishna:2009yz} 
  S.~Gopalakrishna, S.~J.~Lee and J.~D.~Wells,
  %``Dark matter and Higgs boson collider implications of fermions in an abelian-gauged hidden sector,''
  Phys.\ Lett.\ B {\bf 680}, 88 (2009)
  doi:10.1016/j.physletb.2009.08.010
  [arXiv:0904.2007 [hep-ph]].
  %%CITATION = doi:10.1016/j.physletb.2009.08.010;%%
  %28 citations counted in INSPIRE as of 12 Feb 2016

%\cite{Gopalakrishna:2006kr}
\bibitem{Gopalakrishna:2006kr} 
  S.~Gopalakrishna, A.~de Gouvea and W.~Porod,
  %``Right-handed sneutrinos as nonthermal dark matter,''
  JCAP {\bf 0605}, 005 (2006)
  doi:10.1088/1475-7516/2006/05/005
  [hep-ph/0602027].
  %%CITATION = doi:10.1088/1475-7516/2006/05/005;%%
  %53 citations counted in INSPIRE as of 17 Feb 2016
  %
  
   %\cite{Adam:2015rua}
\bibitem{Adam:2015rua} 
  R.~Adam {\it et al.} [Planck Collaboration],
  %``Planck 2015 results. I. Overview of products and scientific results,''
  arXiv:1502.01582 [astro-ph.CO].
  %%CITATION = ARXIV:1502.01582;%%
  %165 citations counted in INSPIRE as of 23 Feb 2016
%

%\cite{Shifman:1978zn}
\bibitem{Shifman:1978zn} 
  M.~A.~Shifman, A.~I.~Vainshtein and V.~I.~Zakharov,
  %``Remarks on Higgs Boson Interactions with Nucleons,''
  Phys.\ Lett.\ B {\bf 78}, 443 (1978).
  doi:10.1016/0370-2693(78)90481-1
  %%CITATION = doi:10.1016/0370-2693(78)90481-1;%%
  %382 citations counted in INSPIRE as of 17 févr. 2016  
  
%\cite{Bertone:2004pz}
\bibitem{Bertone:2004pz} 
  G.~Bertone, D.~Hooper and J.~Silk,
  %``Particle dark matter: Evidence, candidates and constraints,''
  Phys.\ Rept.\  {\bf 405}, 279 (2005)
  doi:10.1016/j.physrep.2004.08.031
  [hep-ph/0404175].
  %%CITATION = doi:10.1016/j.physrep.2004.08.031;%%
  %2253 citations counted in INSPIRE as of 17 févr. 2016

%\cite{Cheng:2012qr}
\bibitem{Cheng:2012qr} 
  H.~Y.~Cheng and C.~W.~Chiang,
  %``Revisiting Scalar and Pseudoscalar Couplings with Nucleons,''
  JHEP {\bf 1207}, 009 (2012)
  doi:10.1007/JHEP07(2012)009
  [arXiv:1202.1292 [hep-ph]].
  %%CITATION = doi:10.1007/JHEP07(2012)009;%%
  %41 citations counted in INSPIRE as of 14 Mar 2016
%

%\cite{Green:2011bv}
\bibitem{SigmaDD.Un} 
  A.~M.~Green,
  %``Astrophysical uncertainties on direct detection experiments,''
  Mod.\ Phys.\ Lett.\ A {\bf 27}, 1230004 (2012)
  doi:10.1142/S0217732312300042
  [arXiv:1112.0524 [astro-ph.CO]];
  %%CITATION = doi:10.1142/S0217732312300042;%%
  %58 citations counted in INSPIRE as of 19 Apr 2016
  %
  %\cite{McCabe:2010zh}
%\bibitem{McCabe:2010zh} 
  C.~McCabe,
  %``The Astrophysical Uncertainties Of Dark Matter Direct Detection Experiments,''
  Phys.\ Rev.\ D {\bf 82}, 023530 (2010)
  doi:10.1103/PhysRevD.82.023530
  [arXiv:1005.0579 [hep-ph]];
  %%CITATION = doi:10.1103/PhysRevD.82.023530;%%
  %111 citations counted in INSPIRE as of 19 Apr 2016dez
  %
  %\cite{Frandsen:2011gi}
%\bibitem{Frandsen:2011gi} 
  M.~T.~Frandsen, F.~Kahlhoefer, C.~McCabe, S.~Sarkar and K.~Schmidt-Hoberg,
  %``Resolving astrophysical uncertainties in dark matter direct detection,''
  JCAP {\bf 1201}, 024 (2012)
  doi:10.1088/1475-7516/2012/01/024
  [arXiv:1111.0292 [hep-ph]].
  %%CITATION = doi:10.1088/1475-7516/2012/01/024;%%
  %107 citations counted in INSPIRE as of 19 Apr 2016dez%
  
\bibitem{DirDetExptCite}
Rick Gaitskell,  Vuk Mandic and Jeff Filippini, http://dmtools.berkeley.edu/limitplots/ .
%

%\cite{Gopalakrishna:2011ef}
\bibitem{Gopalakrishna:2011ef} 
  S.~Gopalakrishna, T.~Mandal, S.~Mitra and R.~Tibrewala,
  %``LHC Signatures of a Vector-like b',''
  Phys.\ Rev.\ D {\bf 84}, 055001 (2011)
  doi:10.1103/PhysRevD.84.055001
  [arXiv:1107.4306 [hep-ph]].
  %%CITATION = doi:10.1103/PhysRevD.84.055001;%%
  %25 citations counted in INSPIRE as of 14 Apr 2016

%\cite{Gopalakrishna:2013hua}
\bibitem{Gopalakrishna:2013hua} 
  S.~Gopalakrishna, T.~Mandal, S.~Mitra and G.~Moreau,
  %``LHC Signatures of Warped-space Vectorlike Quarks,''
  JHEP {\bf 1408}, 079 (2014)
  doi:10.1007/JHEP08(2014)079
  [arXiv:1306.2656 [hep-ph]].
  %%CITATION = doi:10.1007/JHEP08(2014)079;%%
  %30 citations counted in INSPIRE as of 14 Apr 2016  
  %



  
\end{thebibliography}
\end{document}